\documentclass[conference]{IEEEtran}
\usepackage{booktabs}
\usepackage{tikz}
\usepackage{url}
\usepackage{pdfpages}
\usepackage{afterpage}
\usepackage{color, colortbl}
\usepackage{graphicx}
\usepackage{amsmath}
\usepackage{import}
\usepackage{pifont}
\usepackage{algorithm,algpseudocode}
\usepackage{subfigure}
\usepackage{footmisc}
\usepackage{changepage}
\usepackage[bookmarks=false]{hyperref}
\usepackage{makecell}
\usepackage[misc]{ifsym}
\newtheorem{definition}{Definition}
\usepackage[
separate-uncertainty = true,
multi-part-units = repeat
]{siunitx}

\definecolor{Gray}{gray}{0.9}
\definecolor{LightCyan}{rgb}{0.88,1,1}
\pdfoutput=1

\makeatletter
\DeclareRobustCommand*\textsubscript[1]{%
  \@textsubscript{\selectfont#1}}
\def\@textsubscript#1{%
  {\m@th\ensuremath{_{\mbox{\fontsize\sf@size\z@#1}}}}}
\makeatother

\graphicspath{{./figures/}}
\usepackage[flushleft]{threeparttable}
\IEEEoverridecommandlockouts
\begin{document}

\title{A Distributed Synchronous SGD Algorithm with Global Top-$k$ Sparsification for Low Bandwidth Networks}

\author{\IEEEauthorblockN{Shaohuai Shi\IEEEauthorrefmark{2}, Qiang Wang\IEEEauthorrefmark{2}, Kaiyong Zhao\IEEEauthorrefmark{2}, Zhenheng Tang\IEEEauthorrefmark{2}, Yuxin Wang\IEEEauthorrefmark{2}, Xiang Huang\IEEEauthorrefmark{3}, Xiaowen Chu\IEEEauthorrefmark{1}\thanks{*Corresponding author.}\IEEEauthorrefmark{2}\\}
\IEEEauthorblockA{\IEEEauthorrefmark{2}Department of Computer Science, Hong Kong Baptist University
	\\\IEEEauthorrefmark{3}MassGrid.com, Shenzhen District Block Technology Co., Ltd.\\
	\{csshshi, qiangwang, kyzhao, zhtang, yxwang\}@comp.hkbu.edu.hk, xiang.huang@galasports.net, chxw@comp.hkbu.edu.hk}
}

\maketitle

\begin{abstract}
	Distributed synchronous stochastic gradient descent (S-SGD) with data parallelism has been widely used in training large-scale deep neural networks (DNNs), but it typically requires very high communication bandwidth between computational workers (e.g., GPUs) to exchange gradients iteratively. Recently, Top-$k$ sparsification techniques have been proposed to reduce the volume of data to be exchanged among workers and thus alleviate the network pressure. Top-$k$ sparsification can zero-out a significant portion of gradients without impacting the model convergence. However, the sparse gradients should be transferred with their indices, and the irregular indices make the sparse gradients aggregation difficult. Current methods that use AllGather to accumulate the sparse gradients have a communication complexity of $O(kP)$, where $P$ is the number of workers, which is inefficient on low bandwidth networks with a large number of workers. We observe that not all top-$k$ gradients from $P$ workers are needed for the model update, and therefore we propose a novel global Top-$k$ (gTop-$k$) sparsification mechanism to address the difficulty of aggregating sparse gradients. Specifically, we choose global top-$k$ largest absolute values of gradients from $P$ workers, instead of accumulating all local top-$k$ gradients to update the model in each iteration. The gradient aggregation method based on gTop-$k$ sparsification, namely gTopKAllReduce, reduces the communication complexity from $O(kP)$ to $O(k\log P)$. Through extensive experiments on different DNNs, we verify that gTop-$k$ S-SGD has nearly consistent convergence performance with S-SGD, and it has only slight degradations on generalization performance. In terms of scaling efficiency, we evaluate gTop-$k$ on a cluster with 32 GPU machines which are interconnected with 1 Gbps Ethernet. The experimental results show that our method achieves $2.7-12\times$ higher scaling efficiency than S-SGD with dense gradients and $1.1-1.7\times$ improvement than the existing Top-$k$ S-SGD.
\end{abstract}



\section{Introduction} \label{intro}
With the increase of training data volume and the growing complexity of deep neural networks (DNNs), distributed computing environments (such as GPU clusters) are widely adopted to accelerate the training of DNNs. The data-parallel synchronous stochastic gradient descent (S-SGD) method is one of the commonly used optimizers to minimize the objective function of large-scale DNNs \cite{dean2012large}\cite{lecun2015deep}. Compared to SGD on a single worker, S-SGD distributes the workloads to multiple workers to accelerate the training, but it also introduces the communication overhead of exchanging model parameters or gradients in each iteration. Assume that there are $P$ workers training a single DNN model with S-SGD. In every iteration, all workers take different mini-batches of data to calculate the model gradients in parallel. Then they need to average the gradients before updating the model parameters, which involves significant data communications \cite{shi2017performance}. Since the computing power of computational units (e.g., GPUs and TPUs) grows much faster than the growth of network speed, network communication performance has now become the training bottleneck, especially when the communication-to-computation ratio is high \cite{shi2018adag}. Many large IT companies use expensive high-speed networks such as 40/100Gbps IB or Ethernet to alleviate the communication pressure, but still many researchers and small companies can only use consumer-level GPUs connected by low-bandwidth networks such as 1Gig-Ethernet.

\begin{table}[!t]
	\centering
	\begin{threeparttable}
		\caption{Communication complexity of gradient aggregation algorithms}
		\label{table:allreduce}
		\begin{tabular}{|l|c|c|}
			\hline
			Aggregation Algorithm &  Complexity & Time Cost \\\hline
			\hline
			DenseAllReduce & $O(m)$ & $2(P-1)\alpha+2\frac{P-1}{P}m\beta$\\\hline
			TopKAllReduce & $O(kP)$ & $\log(P)\alpha+2(P-1)k\beta$\\\hline
			Ours (gTopKAllReduce)& $O(k\log P)$ & $2 \log(P)\alpha+4k\log(P)\beta$\\\hline
		\end{tabular}
		\begin{tablenotes}
			\item Note: $m$ is the number of model parameters. $P$ is the number of workers. $k=\rho\times m$ is the number of local gradients to be aggregated. $\alpha$ and $\beta$ are system dependent constants.
		\end{tablenotes}
	\end{threeparttable}
\end{table}

To conquer the communication challenge, one can either increase the workload of workers by choosing a large batch size or reduce the required data communications in each iteration. Very recently, many large-batch SGD techniques have been proposed with sophisticated optimization strategies \cite{das2016distributed}\cite{goyal2017accurate}\cite{wang2017stochastic}\cite{you2018imagenet}\cite{jia2018highly} to increase the scaling efficiency without losing the model accuracy. On the other hand, gradient sparsification, quantification and compression methods \cite{hubara2017quantized}\cite{chen2017adacomp}\cite{lin2017deep}\cite{wu2018error}\cite{bernstein2018signsgd}\cite{alistarh2018convergence}\cite{stich2018sparsified} have been proposed to dramatically reduce the size of exchanged gradients without affecting the convergence rate. Among the model/gradient size reduction techniques, the Top-$k$ sparsification is one of the key approaches \cite{aji2017sparse}\cite{lin2017deep}\cite{jiang2018linear} that can sparsify the local gradients to just about $0.001$ density ($99.9\%$ gradients are zeros and there is no need to transfer these zero-out values) \cite{chen2017adacomp}\cite{lin2017deep}.

Top-$k$ sparsification has been a successful gradient compression method with empirical and theoretical studies in \cite{aji2017sparse}\cite{lin2017deep}\cite{stich2018sparsified}, in which researchers have verified that only a small number of gradients are needed to be averaged during the phase of gradient aggregation without impairing model convergence or accuracy. However, the sparsified gradients are generally associated with irregular indices, which makes it a challenge to accumulate the selected gradients from all workers\footnote{Worker and GPU are interchangeable in this paper.} efficiently. The ring-based AllReduce method used on dense gradients (DenseAllReduce) has an $O(P+m)$ communication complexity \cite{chan2007collective}, where $P$ is the number of workers and $m$ is the size of parameters (or gradients). In Top-$k$ sparsification, assume that the density of gradients is $\rho$ on each worker, $k=\rho\times m$, and the corresponding indices of non-zero values are irregular from different workers and iterations; thus it generally needs to transfer $2k$ number of values (gradients and indices) in each iteration. However, with the sparse gradients, the DenseAllReduce method cannot be directly used to accumulate all the sparse gradients with irregular indices, and a recent solution uses the AllGather collective \cite{renggli2018sparcml}, which requires an $O(kP)$ communication complexity. We use TopKAllReduce to denote the method of averaging irregularly indexed top-$k$ gradients by adopting AllGather. When scaling to a large number of workers (i.e., $P$ is large), even high sparsification ratios still generate significant communication overheads.

The main idea of Top-$k$ sparsification is based on the fact that gradients with larger absolute values can contribute more to the model convergence. Theoretical analysis on this idea has been proposed in \cite{wangni2018gradient}\cite{stich2018sparsified}\cite{jiang2018linear}. Therefore, one can further select Top-$k$ gradients from the accumulated results from $P$ groups of top-$k$ values generated by $P$ workers. That is, even though $P$ workers can generate a maximum number of $k\times P$ non-zero gradients after aggregation, the top-$k$ gradients (in terms of absolute values) would be the most important for the model updates. Based on this observation, we propose an efficient Top-$k$ sparsification method to tackle the difficulty of TopKAllReduce with little impacting on the model convergence. Specifically, instead of accumulating the irregularly indexed non-zero gradients from all workers, we choose the global Top-$k$ (gTop-$k$) gradients in terms of absolute values. gTop-$k$\footnote{In this paper, we mainly discuss the decentralized S-SGD with AllReduce to apply gTop-$k$ sparsification. But it is also applicable to the Parameter Server based distributed SGD.} can elegantly make use of a tree structure to select the global top-$k$ values from all workers, which we call gTopKAllReduce, such that the communication complexity is reduced from $O(kP)$ to $O(k\log P)$. We summarize the communication complexities of different gradient aggregation solutions in Table \ref{table:allreduce}.

In this paper, we first implement the gTopKAllReduce algorithm which provides much more efficient global Top-$k$ sparse gradients aggregation from distributed workers. Then we integrate our proposed gTopKAllReduce to gTop-$k$ S-SGD under PyTorch\footnote{\url{https://pytorch.org/}}, which is one of the most popular deep learning frameworks and MPI\footnote{\url{https://www.open-mpi.org/}}. On a 32-node GPU cluster connected by 1-Gbps Ethernet, gTop-$k$ S-SGD achieves 2.7-12.8x speedup than S-SGD with highly optimized libraries Horovod \cite{sergeev2018horovod} and NCCL\footnote{\url{https://developer.nvidia.com/nccl}}. Compared to Top-$k$ S-SGD, gTop-$k$ S-SGD is generally around $1.5$ times faster on the evaluated experiments on various DNNs and datasets. Our contributions are summaries as follows:
\begin{itemize}
	\item We observe that the accumulating results of Top-$k$ sparsification can be further sparsified before being updated to the model.
	\item We propose an efficient global Top-$k$ sparsification algorithm on distributed SGD, called gTop-$k$ S-SGD, to accelerate distributed training of deep neural networks.
	\item We implement the proposed gTop-$k$ S-SGD atop popular framework PyTorch and MPI, and we also release all our experimental parameters for reproducibility\footnote{All experimental settings and source codes can be found at GitHub: \url{https://github.com/hclhkbu/gtopkssgd}}.
	\item gTop-$k$ S-SGD achieves significantly improved scaling efficiency on the real-world applications with various DNNs and datasets under low-bandwidth networks (e.g., 1 Gig-Ethernet).
\end{itemize}

The rest of the paper is organized as follows. We introduce the preliminaries in Section \ref{sec:pre}, in which some background information and the main problem is clarified. In Section \ref{sec:method}, we present our observation from Top-$k$ sparsification and propose an efficient gTop-$k$ S-SGD algorithm. Then we demonstrate the detailed experimental study in Section \ref{sec:eval} and have a discussion in Section \ref{sec:discussion}. Section \ref{sec:related} gives an introduction to the related work, and finally we conclude the paper in Section \ref{sec:conclusion}.

\section{Preliminaries}\label{sec:pre}



\subsection{DNNs}
Deep neural networks (DNNs) are generally stacked with many hierarchical layers, and each layer is a transformer function of the input values. We can formulate an $L$-layer DNN by
\begin{equation}\label{equ:nn}
a^{(l)}=f(W^{(l)},x^{(l)}),
\end{equation}
where $x^{(l)}$ and $a^{(l)}$ are the input and output of layer $l$ ($l=1,2,...,L$) respectively. Inputs of current layer are the outputs of its previous layer(s) (e.g., $x^{l}=a^{(l-1)}$). The function $f$ is the transformer function which consists of an operation (e.g., inner product or convolution) and an activation function (e.g., ReLU). $W^{(l)}$ are the trainable model parameters, which could be iteratively updated during the training process using mini-batch stochastic gradient descent (SGD) optimizers and the backpropagation algorithm.

\subsection{Mini-batch SGD}
The objective function $\mathcal{L}(W,D)$ defines the differences between the prediction values by the DNN and the ground truth. The mini-batch SGD optimizer updates the parameters iteratively to minimize the objective function. To be specific, there are three phases in each iteration during training: 1) Feed-forward phase: a mini-batch of data $D_i$ ($D_i\subset D$) is read as inputs of a DNN, and $D_i$ is fed forward across the neural network from the first layer to the last layer, which finally generates the prediction values to be used by the objective function $\mathcal{L}(W,D)$. 2) Backward-propagation phase: the gradients w.r.t. the parameters and inputs are calculated from the last layer to the first layer. 3) Update phase, the parameters are updated by the afore-generated gradients using the following formula (or its variants):
\begin{equation}
W_{i+1}=W_{i}-\eta\cdot\nabla\mathcal{L}(W_{i},D_{i}),
\end{equation}
where $\eta$ is the learning rate. For a single-worker training, phases 1) and 2) are the main time costs of an iteration, which are computing-intensive tasks. So the average time of one iteration can be approximated by $t_{iter}=t_f+t_b$.

\subsection{Synchronous SGD}
Synchronous SGD (S-SGD) with data parallelism is widely applied to train models with multiple workers (say $P$ workers, and indexed by $g$). Each worker keeps a consistent model at the beginning of each iteration. The workers take different mini-batches of data $D_{i}^{g}$ and forward them by phase 2), and then follow phase 3) to calculate the gradients $\nabla\mathcal{L}(W_{i},D_{i}^{g})$ in parallel. The average gradients from different workers are used to update the model. The update formula of parameters is rewritten as
\begin{equation}\label{equ:ssgd}
W_{i+1}=W_{i}-\eta\frac{1}{P}\sum_{g=1}^{P}G_i^g,
\end{equation}
where $G_i^g=\nabla\mathcal{L}(W_{i},D_{i}^{g})$ denotes the gradients of worker $g$ at the $i^{th}$ iteration. The gradients are located in different workers without shared memory so that the averaging operation of gradients involves communication costs, which could become another system bottleneck. The average iteration time of S-SGD can be approximated by $t_{iter}=t_f+t_b+t_c$. Assume that we use weak-scaling on $P$ workers with S-SGD, the scaling efficiency can be approximated by
\begin{equation}
e=\frac{t_f+t_b}{t_f+t_b+t_c}.
\end{equation}
$t_c$ is generally related to $P$ and the model/gradient size $m$. 

\subsection{DenseAllReduce}
In Eq. \ref{equ:ssgd}, the summation of $G_i^g$ (i.e., $\sum_{g=1}^{P}G_i^g$) can be directly implemented by an AllReduce collective, which is denoted as DenseAllReduce. The ring-based AllReduce algorithm (which is also included in NCCL) is an efficient implementation on the dense-GPU cluster. To understand the time cost of DenseAllReduce, we revisit the time model of the ring-based AllReduce. According to \cite{hoefler2010toward}, the time cost of ring-based AllReduce can be represented by
\begin{equation}\label{equ:allreduce}
t_c^{dar}=2(P-1)\alpha+2\frac{P-1}{P}m\beta,
\end{equation}
where $\alpha$ is the latency (startup time) of a message transfer between two nodes, and $\beta$ is the transmission time per element between two nodes using the alpha-beta model \cite{sarvotham2001connection}. 

\begin{algorithm}[!h]
	\caption{S-SGD with Top-$k$ sparsification on worker $g$ \cite{lin2017deep}\cite{renggli2018sparcml}}
	\label{algo:topkspar} 
	\small
	\textbf{Input: }The dataset: $D$\\
	The initialized weights: $W$ \\
	The mini-batch size per worker: $b$\\
	The number of workers: $P$ \\
	The number of iterations to train: $N$ \\
	The number gradients to select: $k$
	\begin{algorithmic}[1]
		\State $G_0^{g}=0$
		\For{$i=1\rightarrow N$}
		\State Sampling a mini-batch of data $D_i^{g}$ from $D$;
		\State $G_i^{g}=G_{i-1}^{g}+\nabla\mathcal{L}(W_{i},D_{i}^{g})$;
		\State Select threhold $thr=$ the $k^{th}$ largest value of $|G_i^{g}|$;
		\State $Mask=|G_i^{g}|>thr$;
		\State $\widetilde G_i^g=G_i^{g}\odot Mask$; // Mask has $k$ non-zero values
		\State $G_i^k=G_i^{g}\odot \neg Mask$; // Store the residuals
		\State $G_i=$TopKAllReduce($\widetilde G_i^g$); //$G_i=\frac{1}{P}\sum_{g=1}^{P}\widetilde G_i^g$
		\State $W_{i}=W_{i-1}-\eta G_i$;
		\EndFor
		
		\Procedure{TopKAllReduce}{$\widetilde G_i^g$}
		\State $[\mathcal{V}_i^g,\mathcal{I}_{i}^{g}]=\widetilde G_i^g$;
		\State $[\mathcal{V}, \mathcal{I}]=$AllGather($[\mathcal{V}_i^g,\mathcal{I}_{i}^{g}]$);
		\State $G_i=\widetilde G_i^g$;
		\For{$g=0\rightarrow P-1$}
		\State $G_i[\mathcal{I}[g*P:g*(P+1)]]+=V[g*P:g*(P+1)]$;
		\EndFor
		\State $G_i=G_i/P$;
		\State Return $G_i$;
		\EndProcedure
	\end{algorithmic}
\end{algorithm}

\subsection{Top-$k$ sparsification}
From Eq. \ref{equ:allreduce}, it is noted that with $P$ or $m$ becoming large, the communication cost will be linearly increased. To reduce the size of transfer messages $m$, researchers propose Top-$k$ sparsification \cite{lin2017deep} which introduces highly sparse gradients. With Top-$k$ sparsification, each worker only needs to contribute the $k$ largest absolute values of gradients $G_i^g$ to be summed up in each iteration, and the zeroed-out values of gradients are stored locally and accumulated at the next iteration. Both theoretical and empirical studies have verified that the Top-$k$ sparsification has little impact on the model convergence and accuracy \cite{aji2017sparse}\cite{lin2017deep}\cite{stich2018sparsified}. For completeness, the pseudo-code of Top-$k$ sparsification S-SGD is shown in Algorithm \ref{algo:topkspar}. Note that at Line $9$ of Algorithm \ref{algo:topkspar}, the implementation of TopKAllReduce is completely different from the DenseAllReduce for efficiency since the non-zero values of $\widetilde{G_i^g}$ may come from inconsistent indices $\mathcal{I}_{i}^{g}$ from different workers. Efficient implementations of such sparse AllReduce are non-trivial. Current methods \cite{renggli2018sparcml} are using AllGather to implement TopKAllReduce, in which the sparsified gradients are gathered as a dense vector combined with its corresponding indices, say $\widetilde{G}_i^g=[\mathcal{V}_i^g,\mathcal{I}_{i}^{g}]$. Both sizes of $\mathcal{V}_i^g$ and $\mathcal{I}_{i}^{g}$ are $k$. According to the communication model of AllGather \cite{chan2007collective}, the time cost for all-gathering $2k$ values is
\begin{equation}\label{equ:sparallreduce}
	t_c^{tar}=\log(P)\alpha+2(P-1)k\beta.
\end{equation}
From Eq. \ref{equ:sparallreduce}, we can see that with increasing $P$, $t_c^{tar}$ is linearly increased. Therefore, the effect of Top-$k$ sparsification will diminish with the increase of number of workers. In this paper, we propose a global Top-$k$ (gTop-$k$) sparsification algorithm to address this scalability problem.

\section{Methodology}\label{sec:method}
In this section, we first demonstrate some observations from Top-$k$ sparsification S-SGD, and then we present our proposed global Top-$k$ sparsification algorithm. For ease of presentation, we assume that the number of workers $P$ is the power of $2$.

\subsection{Observations from Top-$k$ sparsification}
\begin{figure}[!h]
	\centering
	\includegraphics[width=0.49\linewidth]{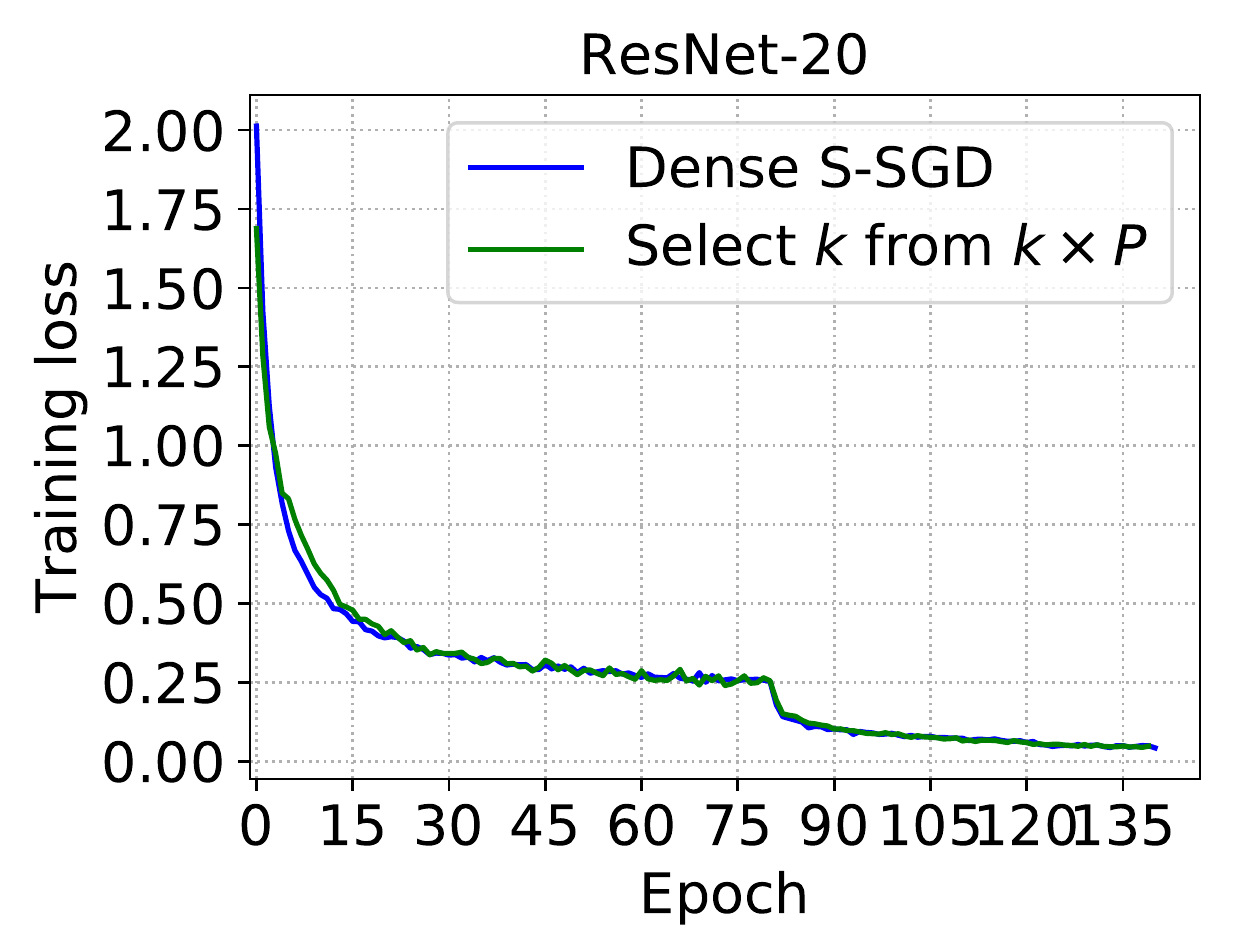}
	\caption{The convergence of ResNet-20 on 4 workers with only $k=0.001\times m$ elements updated at each iteration. The experimental settings can be found in Section \ref{sec:eval}.}
	\label{fig:ideaverify}
\end{figure}

In the previous section, we have introduced Top-$k$ sparsification S-SGD, in which there are $k$ values selected from the local worker and then are accumulated across all the workers. We get insight into the distribution of non-zero values (denoted as $\mathcal{G}_i$) of $G_i$ which is generated by the summation of the sparse gradients from all workers. We found that not all $\mathcal{G}_i$ (whose number of elements is $\mathcal{K}$, and $k\le \mathcal{K}\le k\times P$) contribute to the model convergence. Specifically, $\mathcal{G}_i$ can be further sparsified as $\widetilde\mathcal{G}_i$ such that only a smaller number of non-zero gradients are needed for model updates. In other words, one can further select top-$k$ largest absolute values, $\widetilde\mathcal{G}_i$, from $\mathcal{G}_i$ to update the model while maintaining the model convergence. In this scenario, the selected $\widetilde\mathcal{G}_i$ from $\mathcal{G}_i$ results in the fact that $\mathcal{K}-k$ afore-summed gradients are neither updated to the model nor stored into the local residuals, which could damage the model convergence. Therefore, if only $k$ elements are selected from $\mathcal{G}_i$ to update the model, the remaining $\mathcal{K}-k$ elements should be put back as residuals with corresponding indices so that they can be accumulated locally and eventually contribute to the model update in some future iterations. We empirically verify this idea by training a ResNet DNN, and show the convergence result in Fig. \ref{fig:ideaverify}.

\begin{algorithm}[!h]
	\caption{Naive version S-SGD with gTop-$k$ on worker $g$}
	\label{algo:gtopnaive} 
	\small
	\textbf{Input: }The dataset: $D$\\
	The initialized weights: $W$ \\
	The mini-batch size per worker: $b$\\
	The number of workers: $P$ \\
	The number of iterations to train: $N$ \\
	The number of gradients to select: $k$
	\begin{algorithmic}[1]
		\State $G_0^{g}=0$
		\For{$i=1\rightarrow N$}
		\State Sampling a mini-batch of data $D_i^{g}$ from $D$;
		\State $G_i^{g}=G_{i-1}^{g}+\nabla\mathcal{L}(W_{i},D_{i}^{g})$;
		\State Select threshold $thr=$ the $k^{th}$ largest value of $|G_i^{g}|$;
		\State $Mask=|G_i^{g}|>thr$;
		\State $\widetilde G_i^g=G_i^{g}\odot Mask$; // Mask has $k$ non-zero values
		\State $G_i^g=G_i^{g}\odot \neg Mask$; // Store the residuals
		\State $G_i=$SparseAllReduce($\widetilde G_i^g$); //$G_i=\frac{1}{P}\sum_{g=1}^{P}\widetilde G_i^g$
		\State // At this time all workers have consistent $G_i$
		\State Select global threshold $gThr=$ the $k^{th}$ largest value of $|G_i|$;
		\State $gMask=|G_i|>gThr$;
		\State $\widetilde G_i=G_i\odot gMask$;
		\State $G_i^g+=\widetilde G_i^g\odot \neg gMask \odot Mask$; // Store extra residuals
		\State $W_{i}=W_{i-1}-\eta \widetilde G_i$;
		\EndFor
	\end{algorithmic}
\end{algorithm}

\subsection{The key idea of gTop-$k$}
\begin{figure}[!h]
	\centering
	\includegraphics[width=\linewidth]{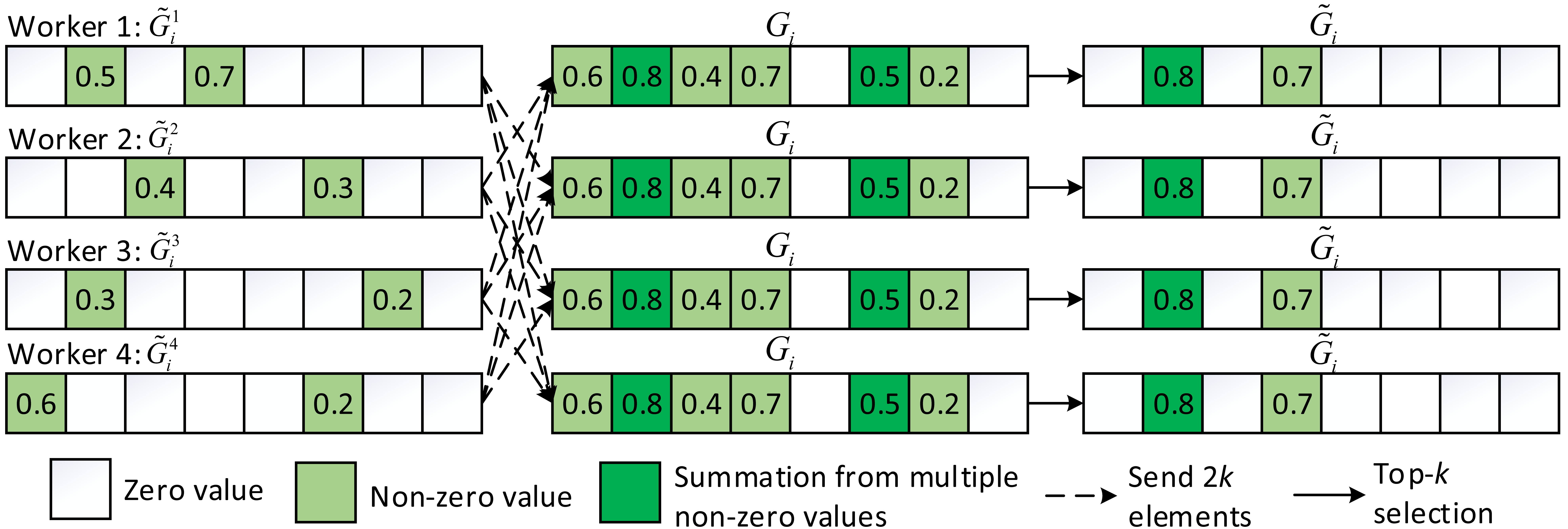}
	\caption{An example of gTop-$k$ using AllGather on $4$ workers, and $k=2$.}
	\label{fig:gtopknaive}
\end{figure}

According to the above observations, it only needs $k$ largest absolute values from all the sparsified gradients $G_i^g$, where $g=1, 2, ..., P$. Therefore, the problem is formulated as the global Top-$k$ (gTop-$k$) selection from $G_i$ instead of using TopKAllReduce, while $\widetilde G_i^g$ are located in distributed workers. We again let $[V_i^g, \mathcal{I}_i^g]$ denote the non-zero values and corresponding indices of $\widetilde G_i^g$ whose number of non-zero values is $k$. We first use the AllGather version to illustrate the key idea of gTop-$k$ sparsification, and then we present our new efficient algorithm for gTop-$k$ sparsification. At Line $10$ of Algorithm \ref{algo:topkspar}, $W_{i}=W_{i-1}-\eta G_i$, $G_i$ with $\mathcal{K}$ non-zero values contributing updates to $W_{i}$. We further sparsify $G_i$ by selecting $k$ largest absolute values from $G_i$. The implementation is shown in Algorithm \ref{algo:gtopnaive}. Please be noted that this version is only used to illustrate the key idea of how to select those gradients to update the model. The efficient algorithm is presented afterwards in the next subsection. An example of gTop-$k$ sparsification using AllGather on $4$ workers is shown in Fig. \ref{fig:gtopknaive}.

\subsection{gTopKAllReduce: An efficient AllReduce algorithm for gTop-$k$ sparsification}
From Eq. \ref{equ:sparallreduce}, we can see that the AllGather collective is inefficient to conduct the AllReduce operation from irregular indexed gradients. Based on the same density, the main purpose of our proposed efficient algorithm is to alleviate the high impact of the variable $P$ on the time cost. For ease of presentation, we first define a Top-$k$ operation, $\top$, of two sparse vectors, say $\widetilde G^{a}$ and $\widetilde G^{b}$, both of which have $k$ non-zero values.

\begin{definition}{A Top-$k$ operation: $\top$.} \label{def:gtopk}
	$\widetilde G^{a,b}=\widetilde G^{a} \top \widetilde G^{b}=mask\odot(\widetilde G^{a}+\widetilde G^{b})$, where $mask=(|\widetilde G^{a}+\widetilde G^{b}|>thr)$, and $thr=$the $k^{th}$ largest value of $|\widetilde G^{a}+\widetilde G^{b}|$.
\end{definition}

\begin{figure}[!h]
	\centering
	\includegraphics[width=0.4\linewidth]{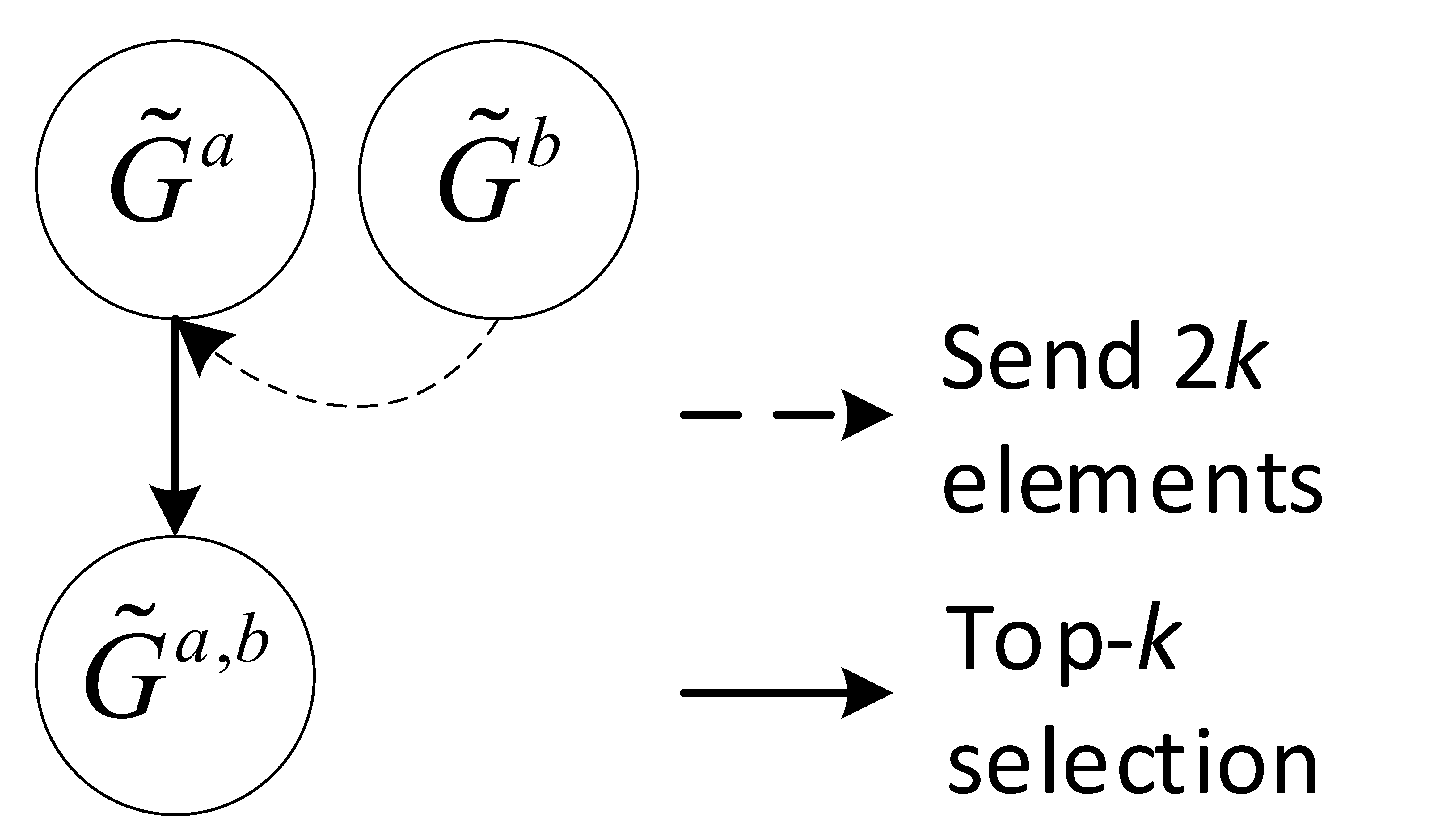}
	\caption{An implementation of $\top$ for two distributed sparse vectors $\widetilde G^{a}$ and $\widetilde G^{b}$. The second worker ($[V^b,\mathcal{I}^b]=\widetilde{G}^b$) with $k$ non-zero elements ($V^b$) combined with $k$ indices ($\mathcal{I}^b$) sends $[V^b,\mathcal{I}^b]$ to the first worker, and then the first worker has the information of indices to add the values received from the second worker, i.e., $\widetilde G^{a}+\widetilde G^{b}$, and the first worker easily computes $\widetilde G^{a,b}=\widetilde G^{a} \top \widetilde G^{b}$ according to Definition \ref{def:gtopk}.}
	\label{fig:gtopk2workers}
\end{figure}

Note that the number of non-zero values of $\widetilde G^{a,b}$ is also $k$. During the training of S-SGD, $\widetilde G^{a}$ and $\widetilde G^{b}$ are located in different workers without shared memory. One should exchange the two sparse vectors to achieve a global Top-$k$ sparse vector: $\widetilde G^{a,b}$. The operation for two distributed workers is shown in Fig. \ref{fig:gtopk2workers}, which demonstrates that $\top$ can be efficiently implemented by a send operation (network communication), followed by a local Top-$k$ selection on a maximum number of $2k$ non-zero values. 

\begin{figure}[!h]
	\centering
	\includegraphics[width=0.7\linewidth]{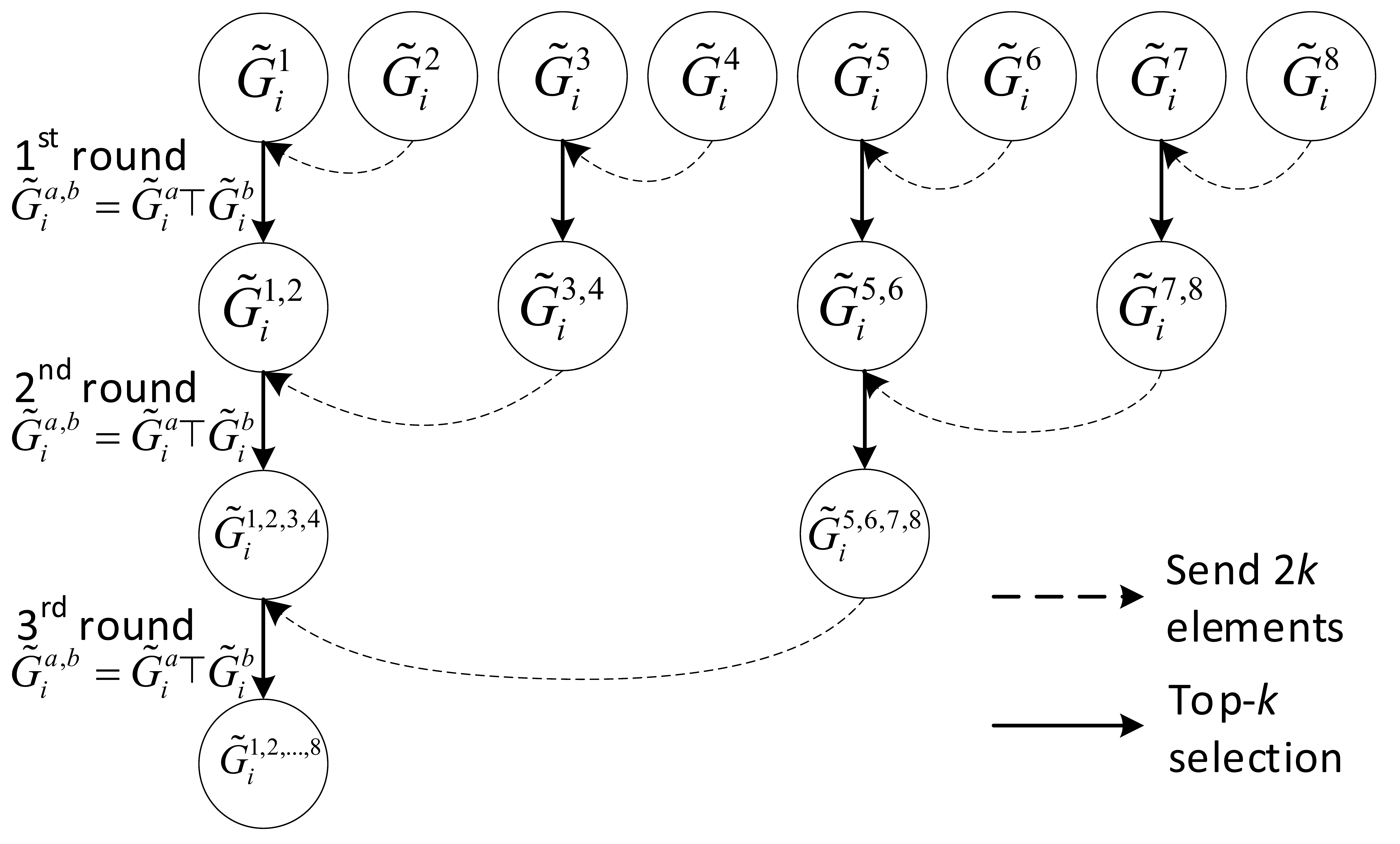}
	\caption{An example of gTop-$k$ for $8$ distributed sparse vectors $\widetilde G^{1}, \widetilde G^{2}, ..., \widetilde G^{8}$. I.e., $\widetilde G = \widetilde G^{1,2,...,8}=\widetilde G^{1}\top\widetilde G^{2}\top...\widetilde G^{8}$. It only requires $log_{2}P=log_{2}8=3$ rounds of network communications to select the global Top-$k$.}
	\label{fig:gtopkefficient}
\end{figure}

When scaling to $P$ workers (assume that $P$ is the power of $2$), since the final $k$ is equal to the local $k$, we propose a tree structure based technique to reduce the total transfer size. To illustrate the tree structure for gTop-$k$, we show an $8$-worker example in selecting the global Top-$k$ values in Fig. \ref{fig:gtopkefficient}. There are $3$ rounds of communications for $8$ workers (i.e., $\log_2 8=3$). At the $j^{th}$ round, there are $\frac{P}{2^j}$ pairs of workers to do the $\top$ operations in parallel. After $3$ rounds, the first worker (rank $0$) finally generates the global Top-$k$ values (i.e., $\widetilde G = \widetilde G^{1,2,...,8}=\widetilde G^{1}\top\widetilde G^{2}\top...\widetilde G^{8}$). 

\begin{algorithm}[!ht]
	\caption{gTopKAllReduce}
	\label{algo:gtopefficient} 
	\small
	\textbf{Input: }The sparsified gradients: $\widetilde G^g$\\
	The number of non-zero elements: $k$ \\
	The number of workers: $P$ \\
	The rank of worker: $g$
	\begin{algorithmic}[1]
		\State $sends=[V^g,\mathcal{I}^g]=\widetilde G^g$;
		\State Initialize $recvs$ with the same as $sends$;
		\State $nRounds=\log P$;
		\For{$i=1\rightarrow nRounds$}
		\State $participateRanks=[1\rightarrow P, step=i]$;
		\If{$g$ in $participateRanks$}
		\State $localRank=participateRanks.index(g)$;
		\If{$localRank\%2==0$}
		\State $source=participateRanks[localRank+1]$;
		\State Recv($recvs$, source=$source$);
		\State $sends=recvs \top sends$;
		\Else
		\State $target=participateRanks[localRank-1]$;
		\State Send($sends$, dest=$target$);
		\EndIf
		\EndIf
		\State Barriar();
		\EndFor
		\State Bcast($sends$, root=0);
		\State $\widetilde G=[V,\mathcal{I}]=sends$;
		\State $Mask=[0]$ in the same number of elements with $\widetilde G$;
		\State $Mask[\mathcal{I}]=1$;
		\State Return $\widetilde G, Mask$;
	\end{algorithmic}
\end{algorithm}

According to the illustration of tree structure based gTop-$k$, we propose the gradients aggregation with gTop-$k$ sparsification, which is called gTopKAllReduce in Algorithm \ref{algo:gtopefficient}. Line $1$ selects the non-zero values from sparse $\widetilde G^g$ to assign the variable ``$sends$'', which should be sent to other workers. Line $2$ allocates the buffer ``$recvs$'' to receive the ``$sends$'' from another worker at each communication round. Lines $4$-$17$ describe the procedure of $\widetilde G = \widetilde G^{1}\top\widetilde G^{2}\top...\widetilde G^{P}$, which is finally stored in ``$sends$'' for the next round communication. The functions ``Recv'' and ``Send'' in Line $10$ and $14$ are a paired operation and can be implemented by MPI. Since the result $\widetilde G$ by far is only stored at the first worker (rank=$0$), Line $19$ broadcasts the $\widetilde G$ to all other workers, which also requires $\log P$ number of communications using the flat-tree algorithm \cite{pjevsivac2007performance}. Finally, Lines $21$ and $22$ record the $Mask$ which indicates the indices that are used in $\widetilde G$.

\subsection{Communication complexity analysis of gTopKAllReduce}
There are two main processes of gTopKAllReduce. The first one is the calculation of $\widetilde G$. From Fig. \ref{fig:gtopkefficient}, the first worker should take part in the communication at every round, so we only need to analyze the communication complexity of the worker with rank $0$. Rank $0$ takes $\log P$ rounds of communications to calculate $\widetilde G$, and it receives $2k$ elements from another worker at each round which takes a time cost of $\alpha+2k\beta$. Thus, the overall time cost of the first process is $\alpha \log P+2k\beta \log P$. In the second process, the global top-$k$ values (i.e., $\widetilde G$) in the first worker should be broadcasted to all the other workers. The broadcast operation takes $\alpha \log P+2k\beta \log P$ according to the flat-tree algorithm. In summary, the time cost of gTopKAllReduce is 
\begin{equation}
t_c^{gar}=2\alpha \log P+4k\beta \log P.
\end{equation}
The communication complexity is much lower than TopKAllReduce especially when $P$ is large.

\subsection{gTop-$k$ S-SGD with gTopKAllReduce}
With the above proposed efficient implementation of gTopKAllReduce, we improve the gTop-$k$ S-SGD in Algorithm \ref{algo:gtopnaive} by replacing Lines $9$-$13$ with a line that invokes gTopKAllReduce shown in Algorithm \ref{algo:gtopefficient}. The improved version of the gTop-$k$ S-SGD training algorithm is presented in Algorithm \ref{algo:gtopkssgd}. Compared to Top-$k$ S-SGD, gTop-$k$ S-SGD only introduces an extra computation (Line $10$ in Algorithm \ref{algo:gtopkssgd}) whose overhead is much smaller than the communication overhead, while gTop-$k$ S-SGD reduces the communication complexity a lot.
\begin{algorithm}[!h]
	\caption{gTopKAllReduce based S-SGD on worker $g$}
	\label{algo:gtopkssgd} 
	\small
	\textbf{Input: }The dataset: $D$\\
	The initialized weights: $W$ \\
	The mini-batch size per worker: $b$\\
	The number of workers: $P$ \\
	The number of iterations to train: $N$ \\
	The number gradients to select: $k$ 
	\begin{algorithmic}[1]
		\State $G_0^{g}=0$
		\For{$i=1\rightarrow N$}
		\State Sampling a mini-batch of data $D_i^{g}$ from $D$;
		\State $G_i^{g}=G_{i-1}^{g}+\nabla\mathcal{L}(W_{i},D_{i}^{g})$;
		\State Select threshold $thr=$ the $k^{th}$ largest value of $|G_i^{g}|$;
		\State $Mask=|G_i^{g}|>thr$;
		\State $\widetilde G_i^g=G_i^{g}\odot Mask$; // Mask has $k$ non-zero values
		\State $G_i^g=G_i^{g}\odot \neg Mask$; // Store the residuals
		\State $\widetilde G_i,gMask=$gTopKAllReduce($\widetilde G_i^g$,$k$,$P$,$g$);
		\State $G_i^g+=\widetilde G_i^g\odot \neg gMask \odot Mask$; // Store extra residuals
		\State $W_{i}=W_{i-1}-\eta \widetilde G_i$;
		\EndFor
	\end{algorithmic}
\end{algorithm}

\section{Experimental Study}\label{sec:eval}
We conduct extensive experiments to evaluate the effectiveness of our proposed gTop-$k$ S-SGD by real-world applications on a 32-GPU cluster. We first validate the convergence of gTop-$k$ S-SGD. Then we evaluate the time cost and efficiency of gTopKAllReduce and compare them with the dense AllReduce (DenseAllReduce) and Top-$k$ AllReduce (gTopKAllReduce) counterparts. After that, we make a comparison on the training efficiency among the three S-SGD algorithms (i.e., S-SGD with dense gradients, Top-$k$ S-SGD, and gTop-$k$ S-SGD). We also break down the training process in an iteration to several phases to analyze the extra overhead introduced by gTop-$k$ sparsification.

\subsection{Experimental setup}
\textbf{Hardware}: The distributed environments are configured as a $32$-node cluster, each with one Nvidia P102-100 GPU. All nodes are connected by a 1-Gbps Ethernet. Details of the hardware are shown in Table \ref{table:hardware}. 

\begin{table}[!ht]
	\centering
	\begin{threeparttable}
		\caption{The experimental setup of hardware.}
		\label{table:hardware}
		\begin{tabular}{|l|l|}
			\hline
			Hardware & Model \\\hline\hline
			CPU & Intel(R) Celeron(R) CPU N3350 @ 1.10GHz \\\hline
			GPU & Nvidia P102-100 (3200 CUDA cores and 5GB Memory)\\\hline
			PCI-e & PCI-e$\times$1 lane with a maximum bandwidth of 250 MB/s  \\\hline
			Memory & 4GB DDR3 with a 16GB swap file\\\hline
			Disk & 256GB SSD \\\hline
			Network & 1 Gbps Ethernet (1GbE) \\\hline
		\end{tabular}
	\end{threeparttable}
\end{table}

\textbf{Software}: All GPU machines are installed with Ubuntu-16.04, Nvidia GPU driver at version 390.48, and CUDA-9.1. The communication libraries are OpenMPI-3.1.1\footnote{\url{https://www.open-mpi.org/}} and NCCL-2.1.5\footnote{\url{https://developer.nvidia.com/nccl}}. We use the highly optimized distributed training library Horovod\footnote{\url{https://github.com/uber/horovod}} \cite{sergeev2018horovod} at version 1.4.1. The deep learning framework is PyTorch at version 0.4.0 with cuDNN-7.1.


\begin{table}[!ht]
	\centering
	\begin{threeparttable}
		\caption{Deep models for training.}
		\label{table:dnns}
\begin{tabular}{|l|l|c|c|c|}
	\hline
	Model & Dataset & $\#$ of Epochs & $b$ & $\eta$ \\\hline\hline
	VGG-16 & Cifar-10 & 140 & 128 & 0.1 \\\hline
	ResNet-20 & Cifar-10 & 140 & 128 & 0.1\\\hline
	AlexNet & ImageNet & 45 & 64 & 0.01 \\\hline
	ResNet-50 & ImageNet & 45 & 256 & 0.01 \\\hline
	LSTM-PTB & PTB & 40 & 100 & 1.0 \\\hline
\end{tabular}
\begin{tablenotes}
	\item Note: All models are trained with 32-bit floating points.
\end{tablenotes}
	\end{threeparttable}
\end{table}

\textbf{DNNs}: We choose various DNNs from several areas of AI applications with different data sets. The data sets include Cifar-10 \cite{krizhevsky2010cifar} that contains $50,000$ training samples, ImageNet \cite{deng2009imagenet} that contains about $1.2$ million samples for image classification, and the Penn Treebank corpus (PTB) \cite{marcus1993building} that contains $923,000$ training samples for language modeling. For the Cifar-10 data set, we use the VGG-16 model \cite{simonyan2014very} and the ResNet-20 model \cite{he2016deep}. For the ImageNet data set, the AlexNet model \cite{krizhevsky2014one} and the ResNet-50 model \cite{he2016deep} are used. We exploit a 2-layer LSTM language model (LSTM-PTB) for the PTB dataset, which is similar as in \cite{lin2017deep}. The details of the deep learning models are given in Table \ref{table:dnns}. We use momentum SGD with a momentum of $0.9$ to train all models.

\subsection{Convergence comparison}
The convergence of Top-$k$ sparsification S-SGD has been verified to be nearly consistent with the dense version in much previous work \cite{aji2017sparse}\cite{lin2017deep}\cite{stich2018sparsified}, so we would not include the convergence curves of Top-$k$ S-SGD. We compare our gTop-$k$ S-SGD with the original S-SGD with dense gradients running on $4$ workers. It has been shown that the warmup strategy in the first several epochs helps the model converge better \cite{lin2017deep}, so we adopt a similar warmup configuration. To be specific, the first $4$ epochs use the dynamic densities of $[0.25, 0.0725, 0.015, 0.004]$ and small learning rates, and the remaining epochs adopt a density of $0.001$ (for CNNs) or $0.005$ (for LSTM).

\begin{figure}[!h]
	\centering
	\subfigure
	{
		\includegraphics[width=0.49\linewidth]{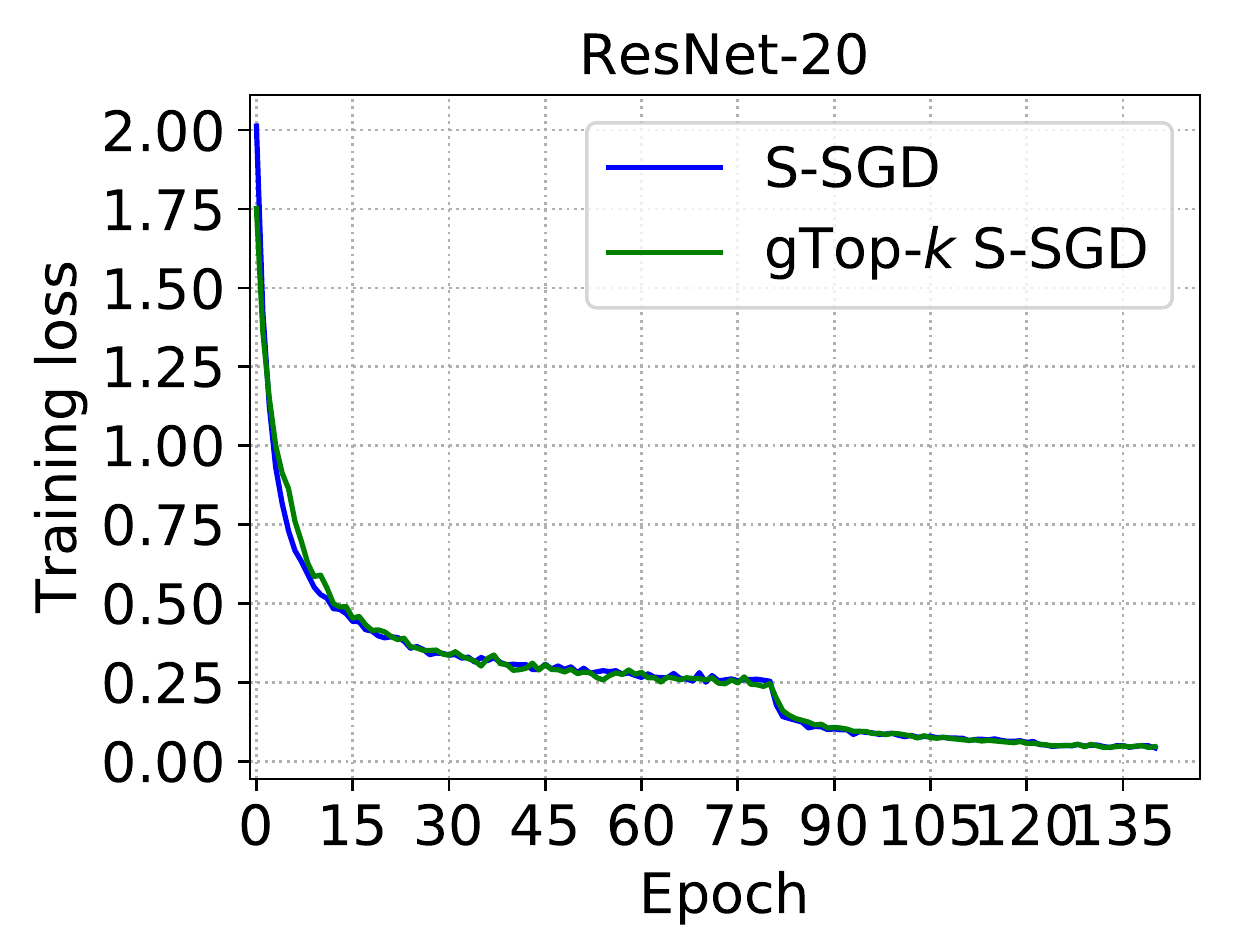}
	}\hspace{-5mm}
	\subfigure
	{
		\includegraphics[width=0.49\linewidth]{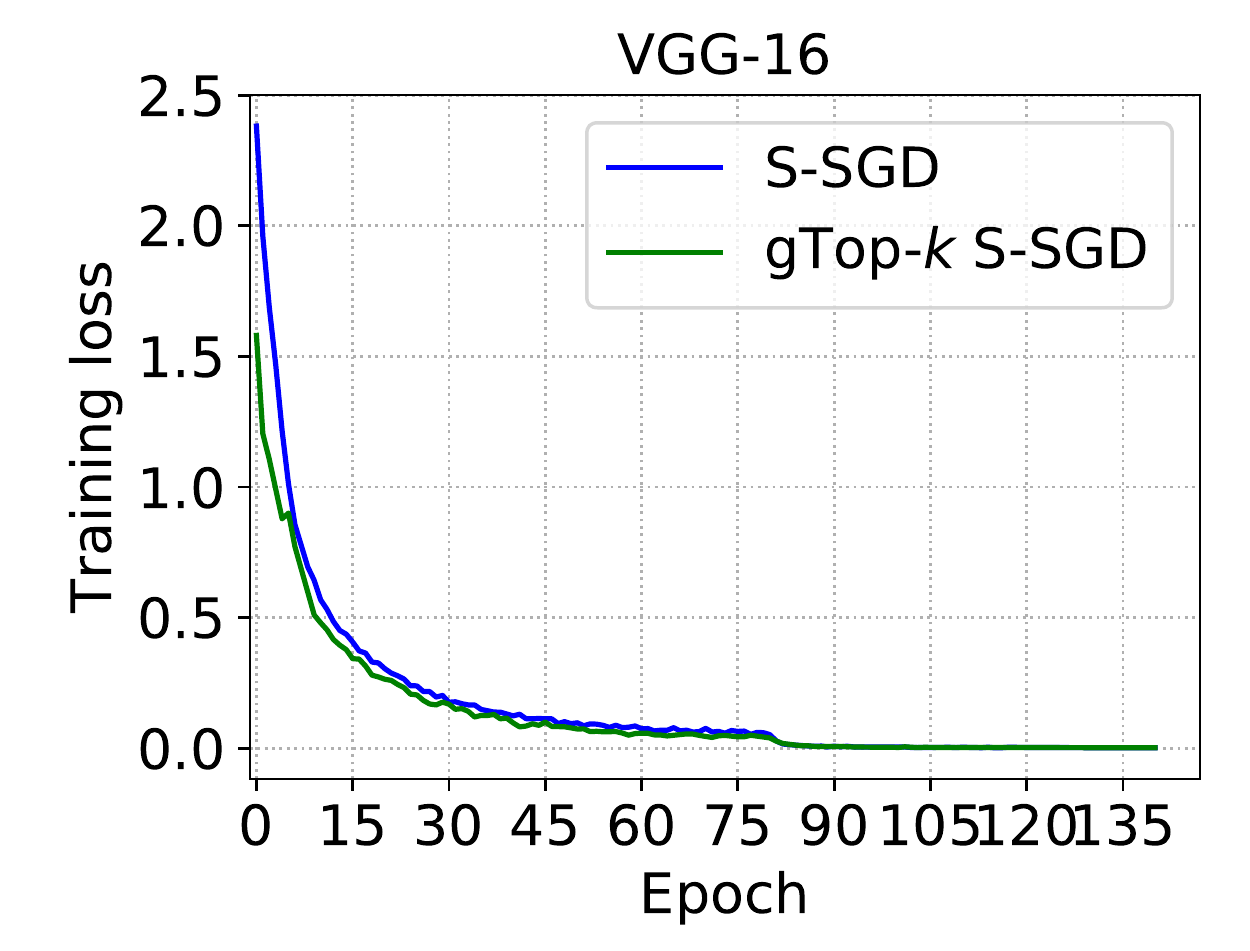}
	}
	\caption{The convergences of VGG-16 and ResNet-20 with $P=4$.}
	\label{fig:convcifar10}
\end{figure}
\textbf{Convergence on the Cifar-10 data set}: The convergences of VGG-16 and ResNet-20 models are shown in Fig. \ref{fig:convcifar10}. The results show that the convergence rate of ResNet-20 is almost the same as the baseline, while the VGG-16 model even converges slightly better than the baseline.

\begin{figure}[!h]
	\centering
	\subfigure
	{
		\includegraphics[width=0.49\linewidth]{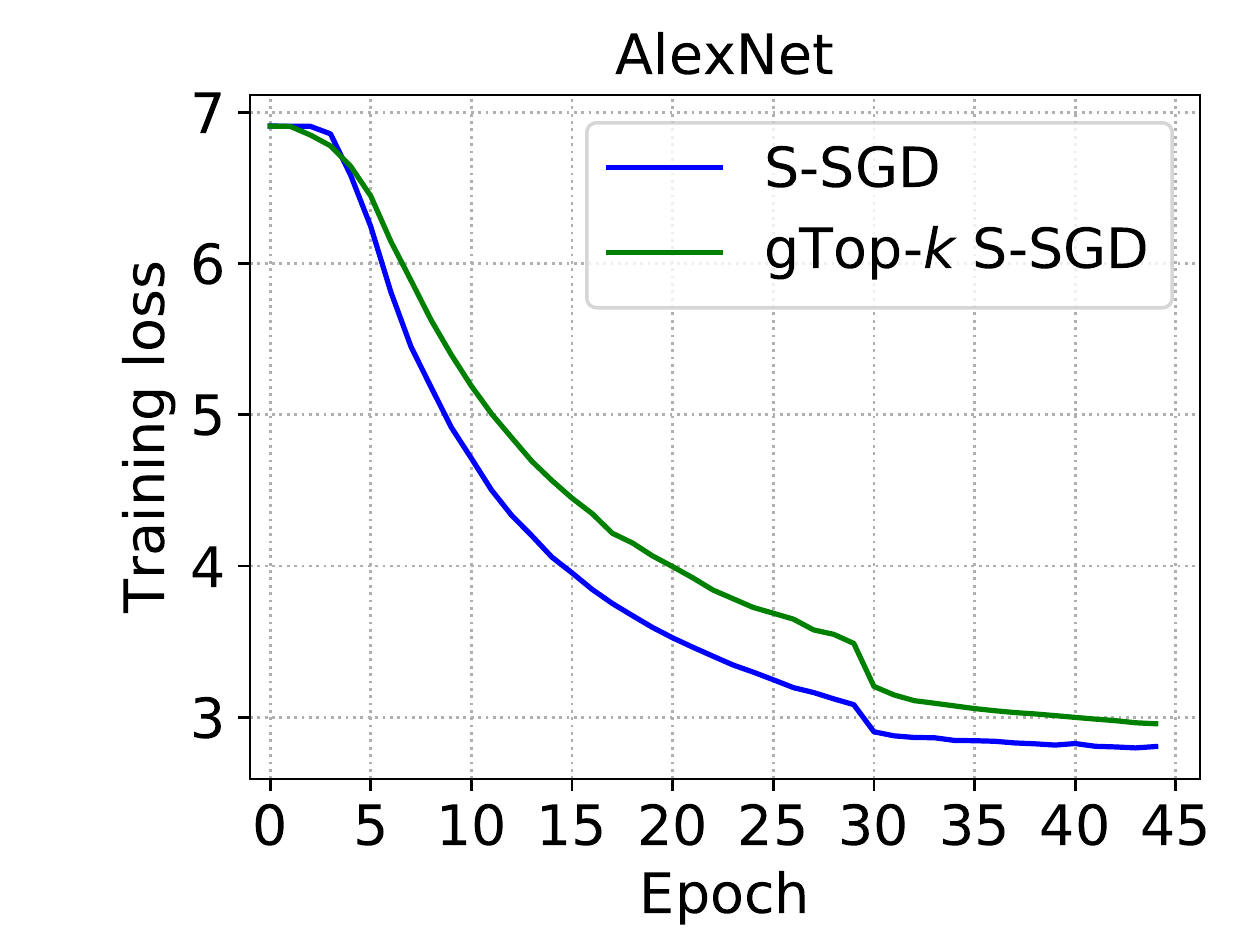}
	}\hspace{-5mm}
	\subfigure
	{
		\includegraphics[width=0.49\linewidth]{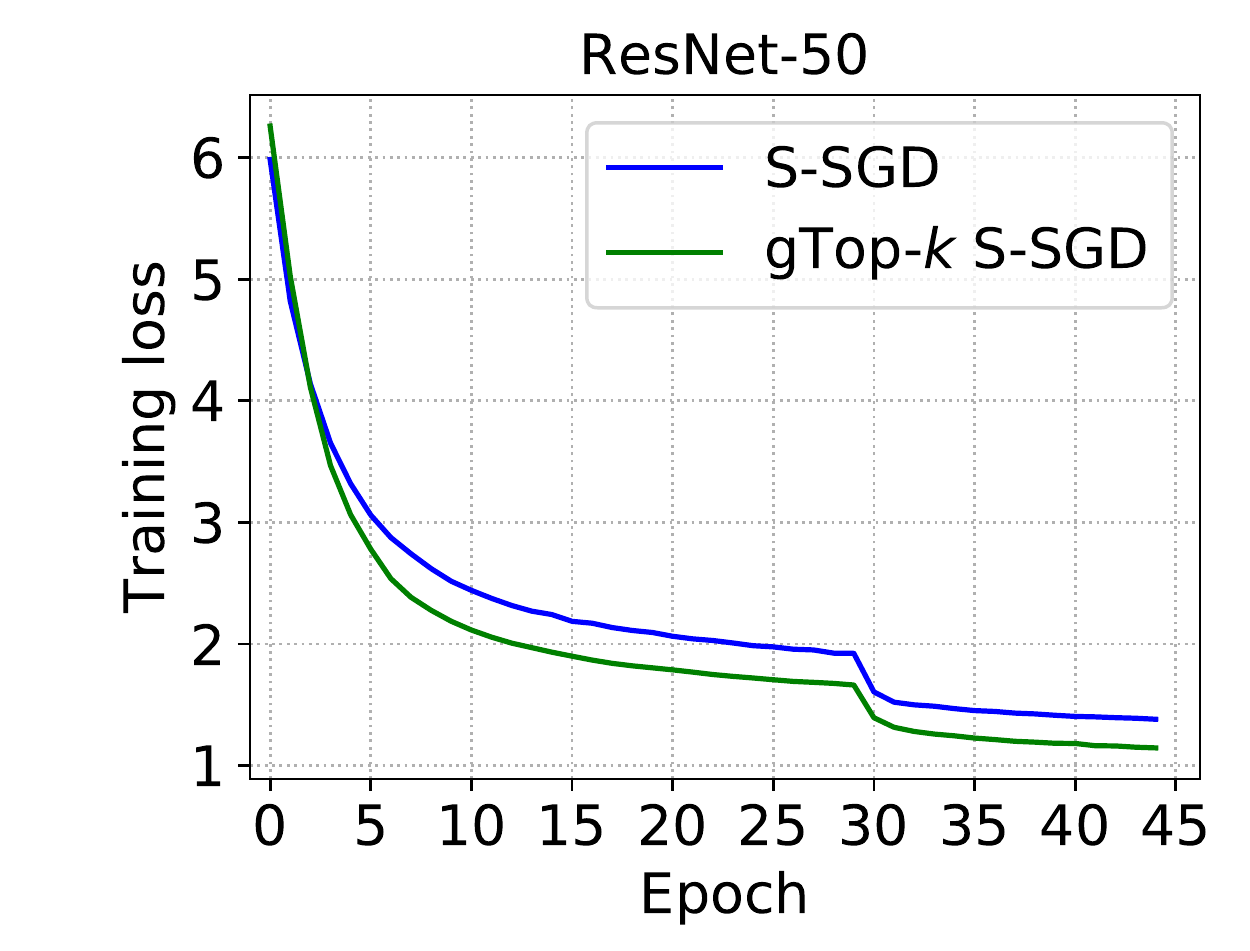}
	}
	\vspace{-15pt}
	\caption{The convergences of AlexNet and ResNet-50 with $P=4$.}
	\label{fig:convimagenet}
\end{figure}
\textbf{Convergence on the ImageNet data set}: The convergences of AlexNet and ResNet-50 models are shown in Fig. \ref{fig:convimagenet}. The results show that the convergence rates of the two CNNs are close to the baselines. On the AlexNet model, the convergence of gTop-$k$ S-SGD with $\rho=0.001$ is slightly worse than the baseline, which could be caused by unbalanced parameters between convolutional layers and fully connected layers with the same low density. On the other hand, gTop-$k$ sparsification works very well on the ResNet-50 model, which converges even faster than the baseline.

\begin{figure}[!h]
	\centering
	\subfigure
	{
		\includegraphics[width=0.49\linewidth]{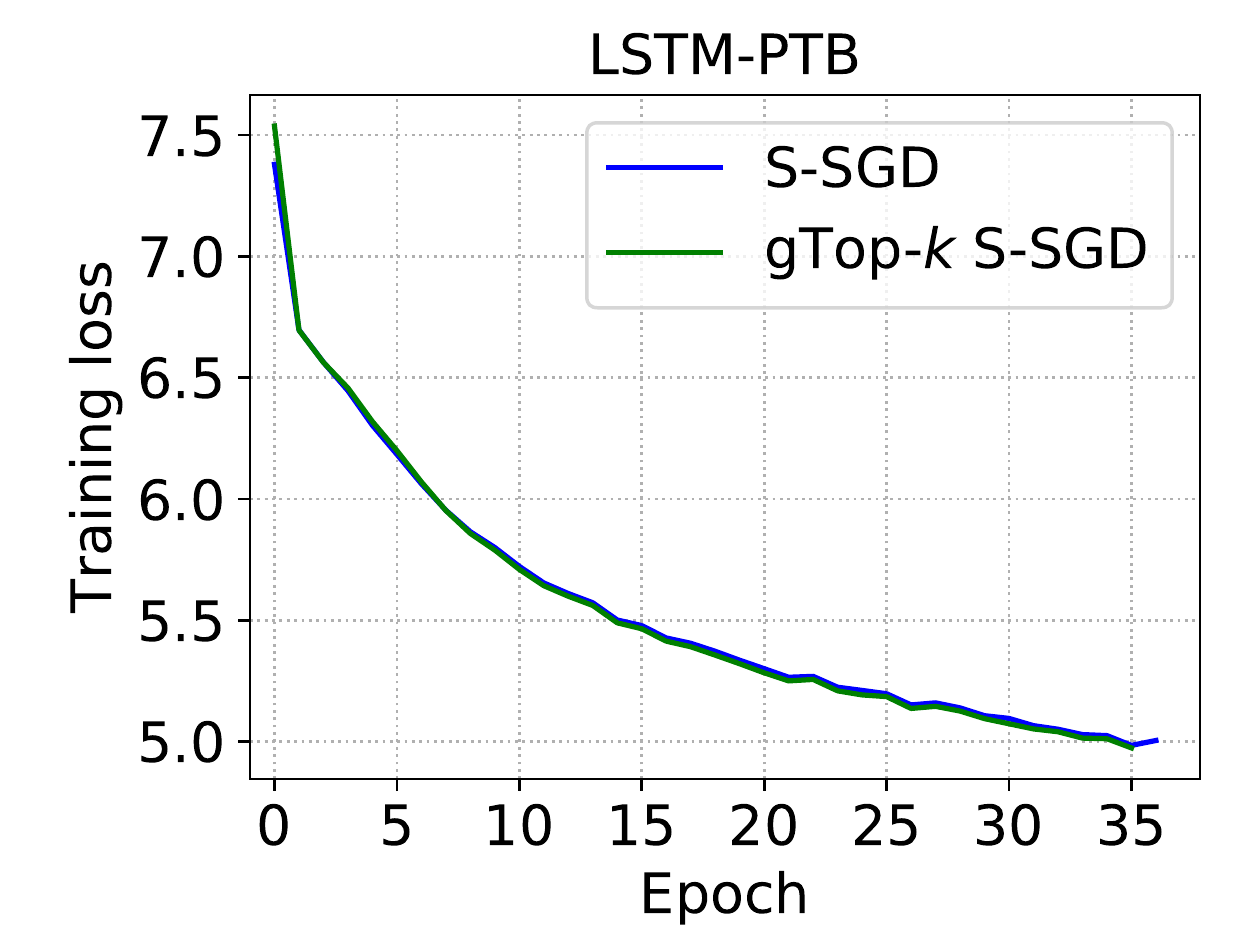}
	}\hspace{-5mm}
	\caption{The convergence of LSTM with $P=4$ and $\rho=0.005$.}
	\label{fig:convlstm}
\end{figure}

\textbf{Convergence on the LSTM network}: The convergence of LSTM-PTB on the PTB data set is shown in Fig. \ref{fig:convlstm}. It is noted that the convergence of gTop-$k$ S-SGD is almost the same as that of S-SGD under a density of $0.005$.

In summary, three different types of DNNs from different data sets show that our proposed gTop-$k$ S-SGD would not damage the model during training and keeps very close model convergence to dense S-SGD.

\subsection{Communication speed}
To set the baseline, we first test the point-to-point communication performance with various sizes of messages because the performance of point-to-point communication plays an essential role in MPI collectives. Then we compare the communication performance of TopKAllReduce and gTopKAllReduce in different sizes of sparse vectors according to Table \ref{table:allreduce}.

\begin{figure}[!h]
	\centering
	\includegraphics[width=0.5\linewidth]{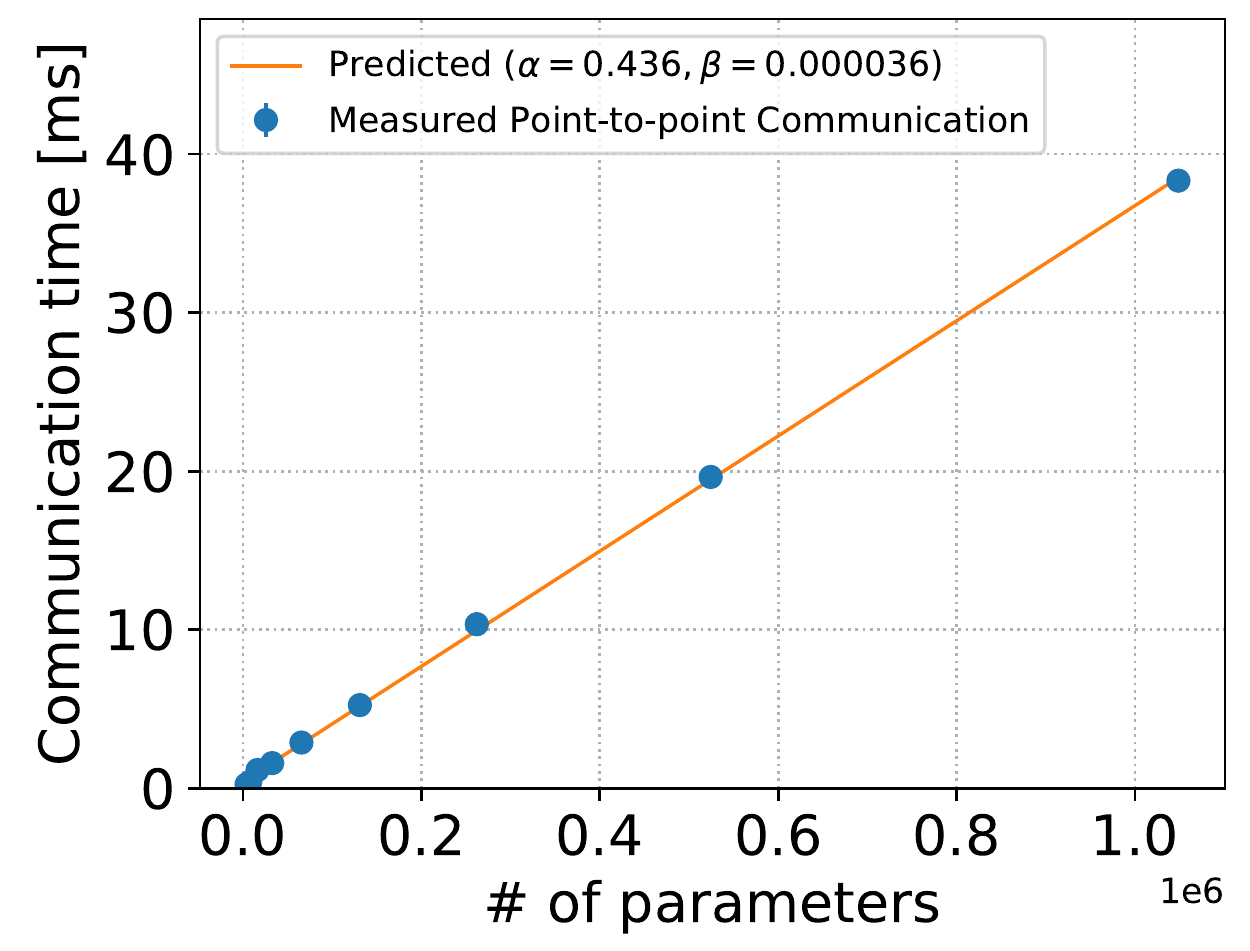}
	\caption{Data transfer time in milliseconds with respective to the size of message on our experiment cluster.}
	\label{fig:p2platency}
\end{figure}

\textbf{Point-to-point communication}: We test the point-to-point communication speed by using OSU Micro-Benchmark\footnote{\url{http://mvapich.cse.ohio-state.edu/benchmarks/}} at the version $5.5$. The time costs of the point-to-point communication between two machines are shown in Fig. \ref{fig:p2platency}, in which we run $5$ times to calculate the mean and standard variance from the reported values. It can be seen that the time used for transferring a message is a linear function with the size of the message, which verifies the $\alpha$-$\beta$ model. Based on the measured data, we can get $\alpha=0.436ms$ and $\beta=3.6\times 10^{-5}ms$.

\begin{figure}[!h]
	\centering
	\subfigure
	{
		\includegraphics[width=0.49\linewidth]{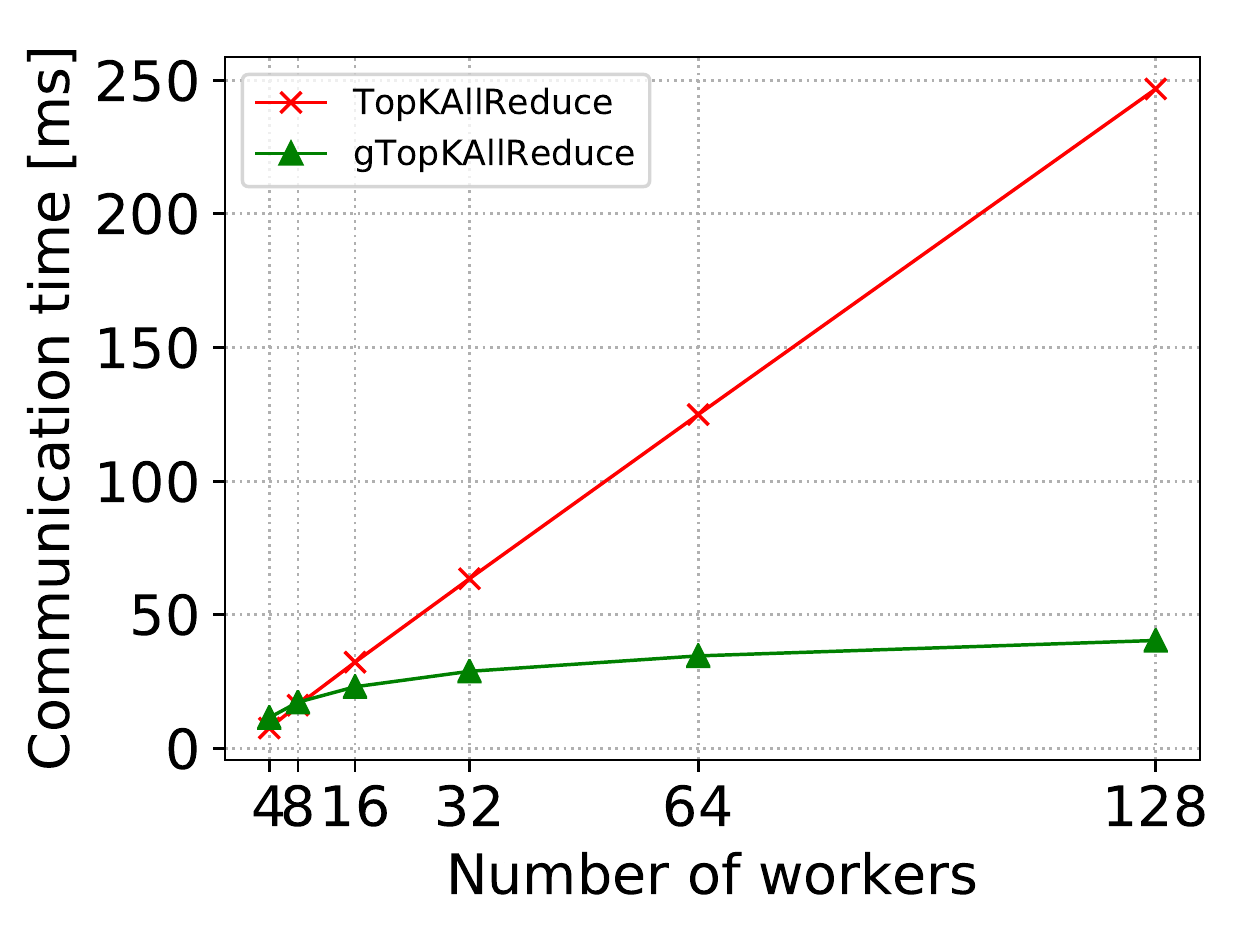}
	}\hspace{-5mm}
	\subfigure
	{
		\includegraphics[width=0.49\linewidth]{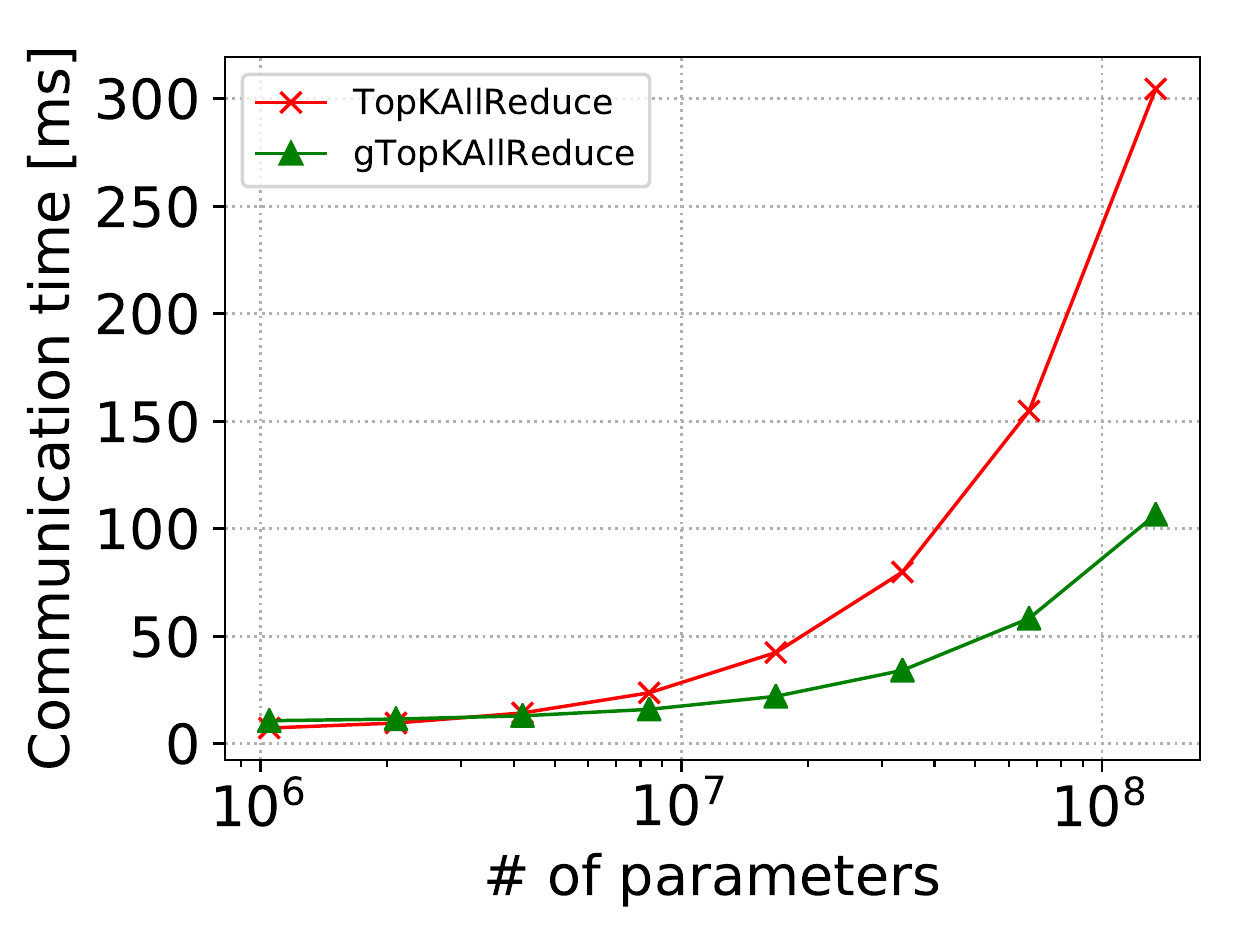}
	}
	\vspace{-15pt}
	\caption{Left: Time used for AllReduce algorithms on different number of workers with $m=25\times 10^6$ parameters, and the density of $\rho=0.001$. Right: The time cost with respective to the number of parameters on 32 workers.}
	\label{fig:arspeed}
\end{figure}

\textbf{Time performance of AllReduce operations}: Since $P$ and $m$ are two main factors affecting the performance of TopKAllReduce and gTopkAllReduce, we compare their time performances in two dimensions (i.e., $P$ and $m$) based on the measured $\alpha$, $\beta$ and the time cost models in Table. \ref{table:allreduce}. First, we compare the time cost with different number of workers (from 4 to 128) based on $m=25\times 10^6$ (the approximate model size of ResNet-50) and $\rho=0.001$. Second, in the configuration of a cluster with 32 workers, we make a comparison on how the time cost changes with the size of parameters increases. The time comparison is shown in Fig. \ref{fig:arspeed}. From the left of Fig. \ref{fig:arspeed}, when the number of nodes is small, TopKAllReduce is slightly faster than gTopKAllReduce. However, when the number of nodes increases to $16$, TopKAllReduce becomes much worse than gTopKAllReduce. Furthermore, our proposed gTopKAllReduce is much more efficient than TopKAllReduce when scaling to large sizes of messages. To summarize, a larger number of workers or a larger message size would make gTopKAllReduce more efficient than TopKAllReduce.

\subsection{Scaling efficiency}

\begin{figure}[!h]
	\centering
	\subfigure
	{
		\includegraphics[width=0.49\linewidth]{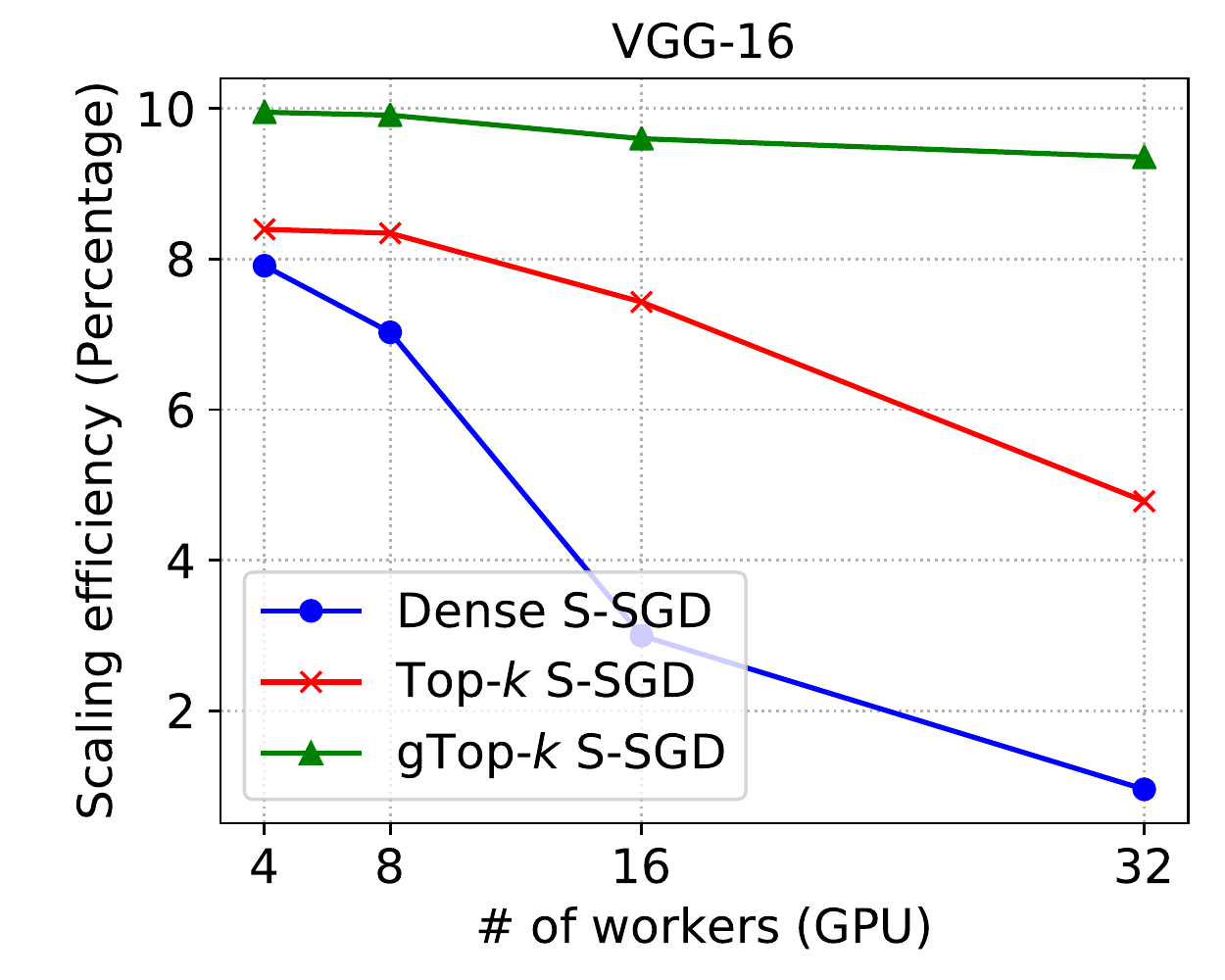}
	}\hspace{-5mm}
	\subfigure
	{
		\includegraphics[width=0.49\linewidth]{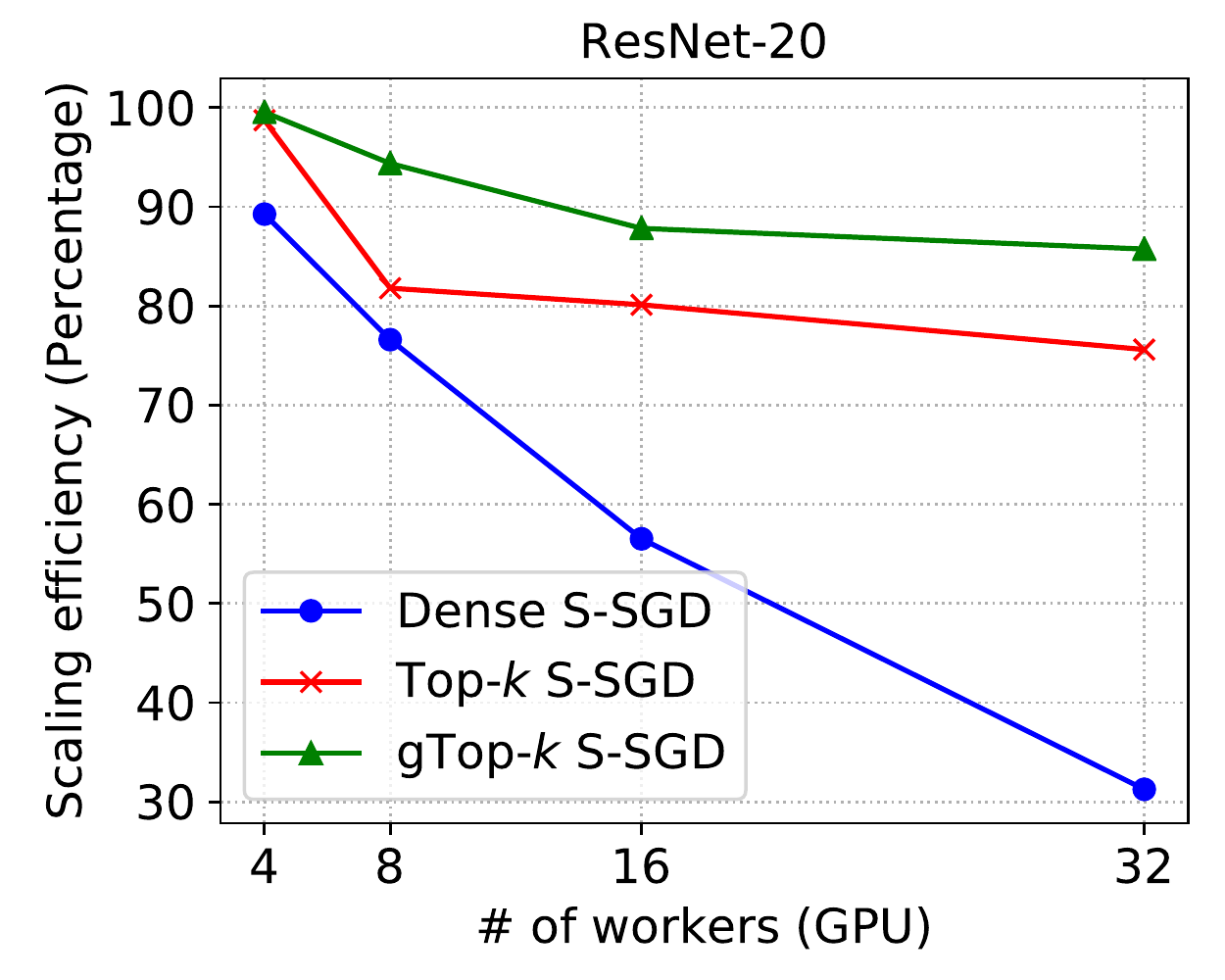}
	}
	\subfigure
	{
		\includegraphics[width=0.49\linewidth]{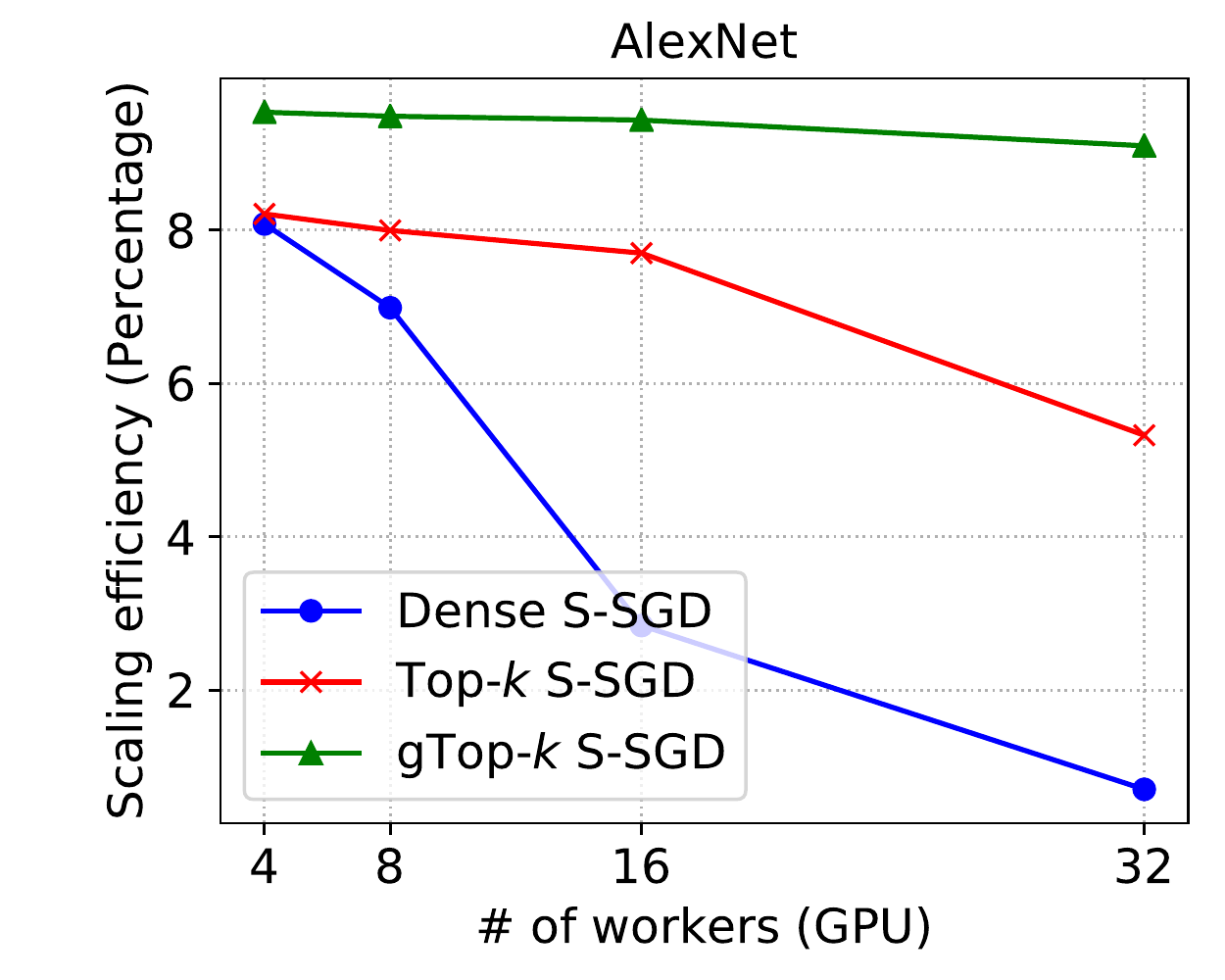}
	}\hspace{-5mm}
	\subfigure
	{
		\includegraphics[width=0.49\linewidth]{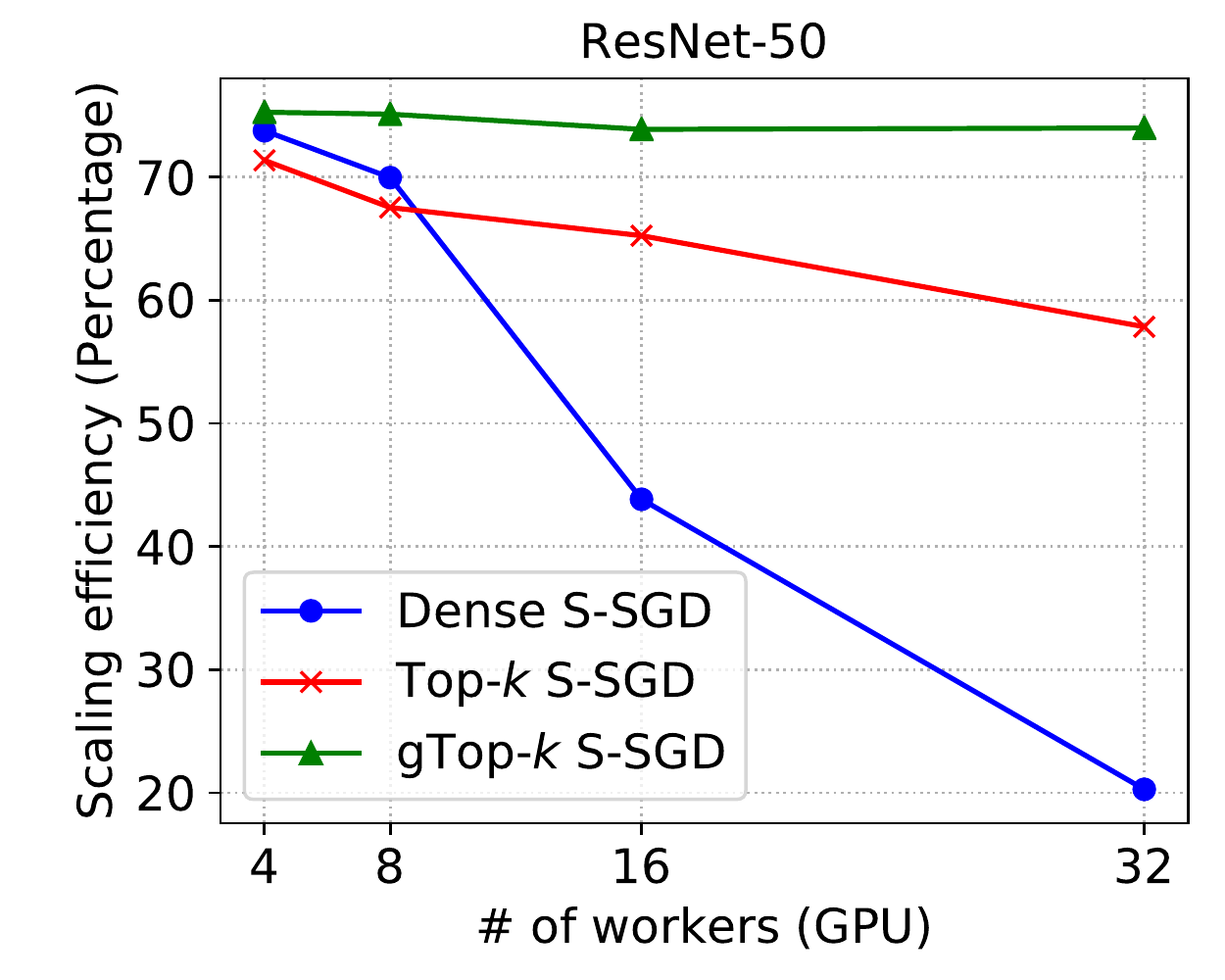}
	}
	\caption{Comparison of scaling efficiency of S-SGD with dense AllReduce (DenseAllReduce), Top-$k$ sparsification (TopKAllReduce) and gTop-$k$ sparsification (gTopKAllReduce), where $k=0.001\times m$. The higher the better.}
	\label{fig:scaling}
\end{figure}

The scaling efficiency of S-SGD with three different AllReduce algorithms are shown in Fig. \ref{fig:scaling}. It can be seen that the dense S-SGD has worst scaling efficiency because the full size of gradients makes the communication very slow on 1GbE clusters. The Top-$k$ S-SGD achieves some improvement on a small number of workers than S-SGD, but it has an obvious performance decrease when scaling to $32$ GPUs. However, our proposed algorithm gTop-$k$ S-SGD achieves much more stable scaling efficiency even on clusters with a larger number of GPUs. For example, when scaling to $32$ GPUs, our proposed gTop-$k$ S-SGD achieves $6.7\times$ faster than dense S-SGD on average, and $1.4 \times$ improvement on average compared to Top-$k$. Particularly, gTop-$k$ S-SGD is up to $12\times$ and $1.7\times$ than S-SGD and Top-$k$ S-SGD respectively on the AlexNet model. The performance improvement will increase with the increase of number of workers.
Summary of the training throughput on different models is shown in Table. \ref{table:gtopkspeedup}. 

\begin{table}[!ht]
	\centering
	\begin{threeparttable}
		\caption{The system training throughput on a 32-GPU cluster.}
		\label{table:gtopkspeedup}
		\begin{tabular}{|l|c|c|c|c|c|}
			\hline
			Model & Dense S-SGD & Top-$k$ & gTop-$k$ & $g/d$ & $g/t$ \\\hline\hline
			VGG-16 & 403 & 2016 & 3020 & 7.5$\times$ & 1.5$\times$ \\\hline
			ResNet-20 & 9212 & 22272 & 25280 & 2.7$\times$ & 1.1$\times$\\\hline
			AlexNet & 39&296& 505 & 12.8$\times$ &  1.7$\times$ \\\hline
			ResNet-50 & 343 & 978 & 1251 & 3.65$\times$ & 1.3$\times$\\\hline
		\end{tabular}
		\begin{tablenotes}
			\item Note: The throughput is measured with processed images per second. $g/d$ indicates the speedup of gTop-$k$ compared to the dense one, and $g/t$ indicates the speedup of gTop-$k$ compared to Top-$k$.
		\end{tablenotes}
	\end{threeparttable}
\end{table} 

\subsection{Time performance analysis}
\begin{figure}[!h]
	\centering
	\includegraphics[width=0.5\linewidth]{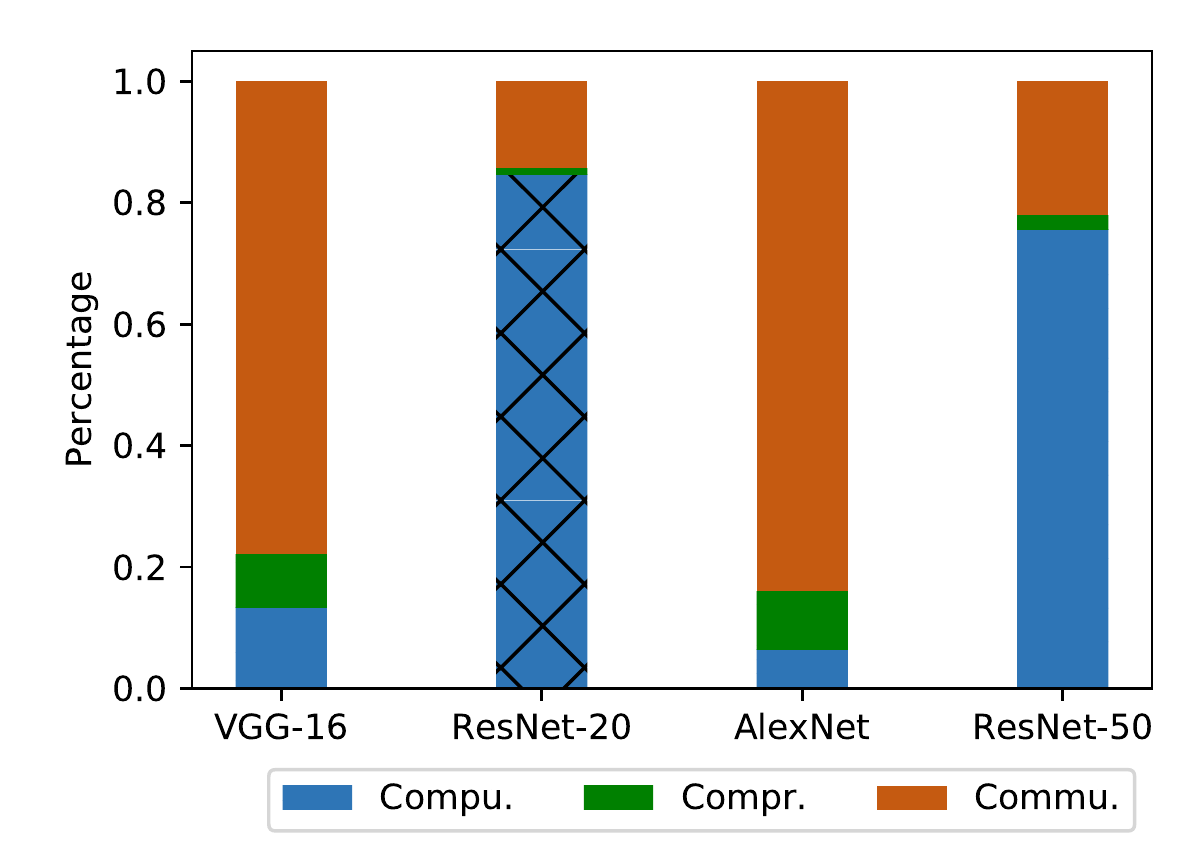}
	\caption{Time breakdown of computation, compression and communication. ``Compu.'' indicates forward and backward computation, ``Compr.'' indicates the compression (sparsification) operation, and ``Commu.' indicates gTop-$k$ gradients communication.}
	\label{fig:breakdown}
\end{figure}

We use the cases of $32$ workers to analyze the time performance of gTop-$k$ S-SGD. We break down the time of an iteration into three parts: GPU computation time ($t_{compu.}$), local sparsification time ($t_{compr.}$), and communication time ($t_{commu.}$). The results are shown in Fig. \ref{fig:breakdown}. On one hand, in the time breakdown of VGG-16 and AlexNet models, the communication overheads are much larger than computation because VGG-16 and AlexNet have three fully connected layers with a large number of parameters, while the computation is relatively fast. These also reflect that the scaling efficiency is low in Fig. \ref{fig:scaling} of S-SGD even with gTop-$k$ sparsification. On the other hand, the communication and sparsification overheads are much smaller than the computation with ResNet-20 and ResNet-50, which indicates low communication-to-computation ratios, so that the scaling efficiency can be up to $80\%$ even on the low-bandwidth network. Furthermore, it is noted that the time used by gradient sparsification is comparable to the computation time on VGG-16 and AlexNet models. The main reason is that Top-$k$ selection on GPU is inefficient and it could be non-trivial to be highly parallelized on SIMD architectures \cite{mei2014benchmarking}\cite{shanbhag2018efficient}. We will leave this as our future optimization direction.

\section{Discussion}\label{sec:discussion}

\subsection{Convergence sensibility to the density}
\begin{figure}[!h]
	\centering
	\subfigure
	{
		\includegraphics[width=0.49\linewidth]{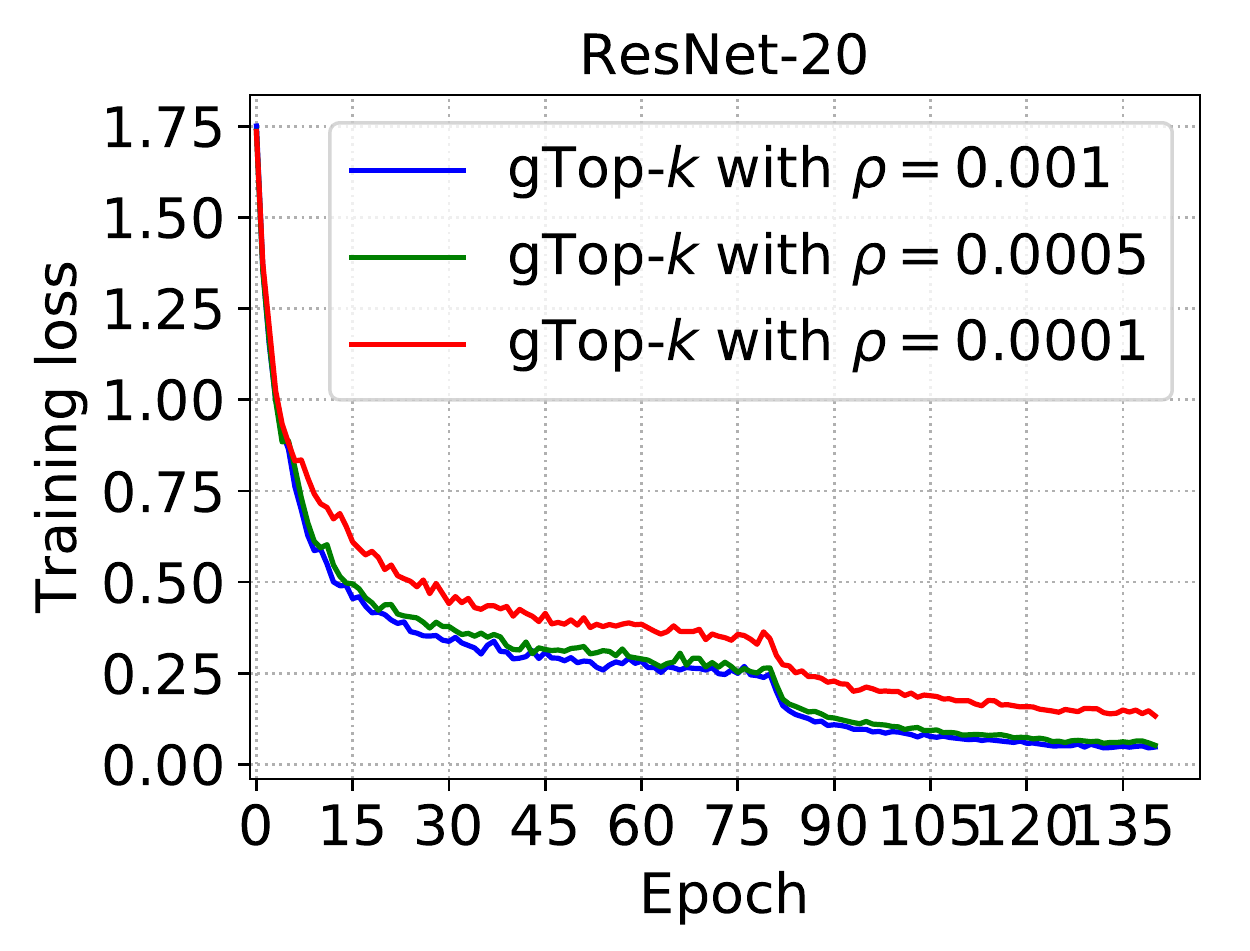}
	}\hspace{-5mm}
	\subfigure
	{
		\includegraphics[width=0.49\linewidth]{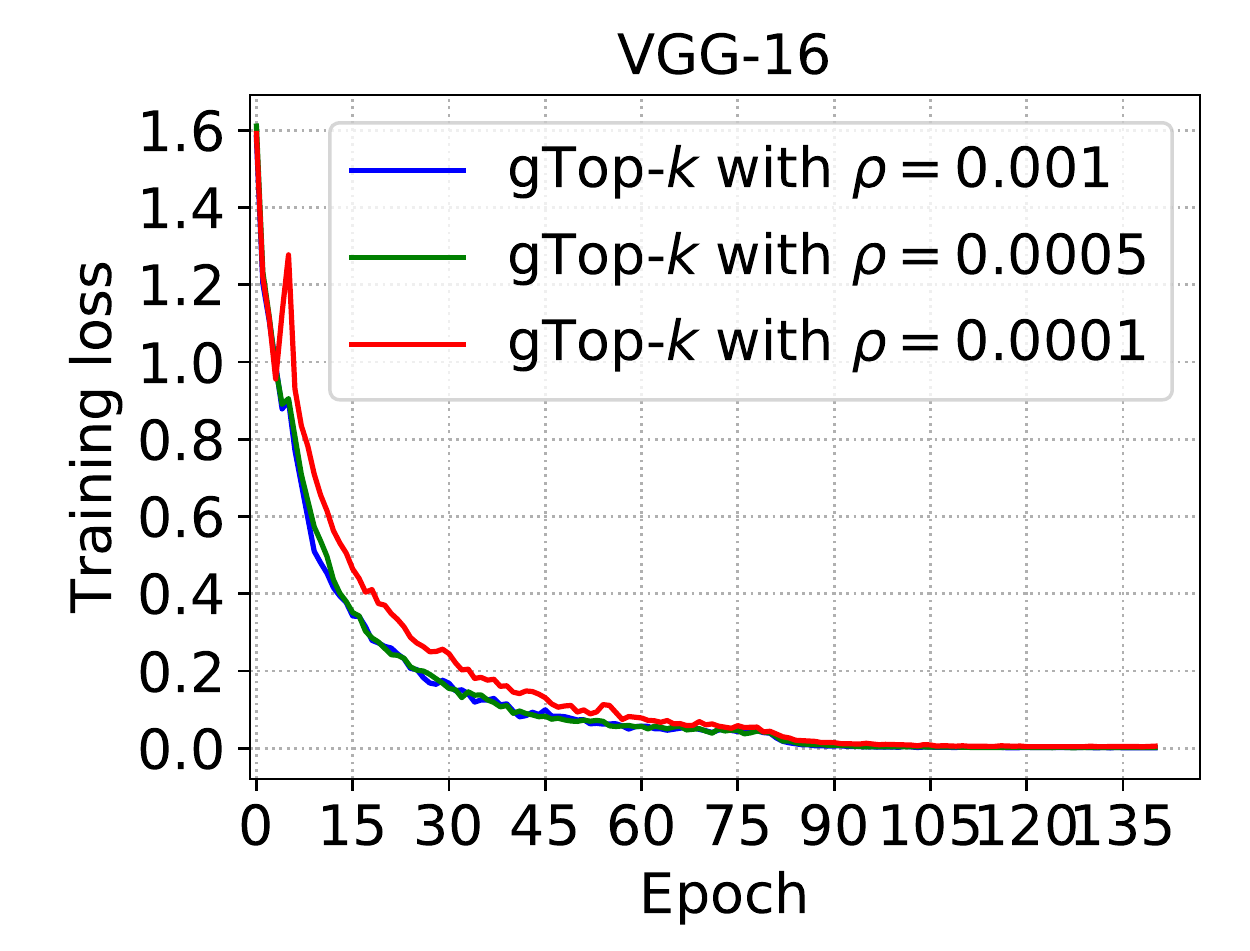}
	}
	\caption{Convergences with different $\rho$ on 4 workers.}
	\label{fig:sense}
\end{figure}

To understand the sensibility of the convergence to the density, we run the experiments with different values of the density $\rho$ using VGG-16 and ResNet-20 on the Cifar-10 data set on 4 workers. The convergence curves are shown in Fig. \ref{fig:sense}. It can be seen that even a very low density of $0.0005$ does not have a big impact on the model convergence to both models. However, a trade-off should be made to balance the high sparsification ratio and the convergence speed. One one hand, the higher sparsification would bring higher scaling efficiency to a larger number of workers. On the other hand, one should also be careful to the upper bound of the sparsity that would hurt the model convergence.

\subsection{Convergence sensibility to the mini-batch size}
\begin{figure}[!h]
	\centering
	\subfigure
	{
		\includegraphics[width=0.49\linewidth]{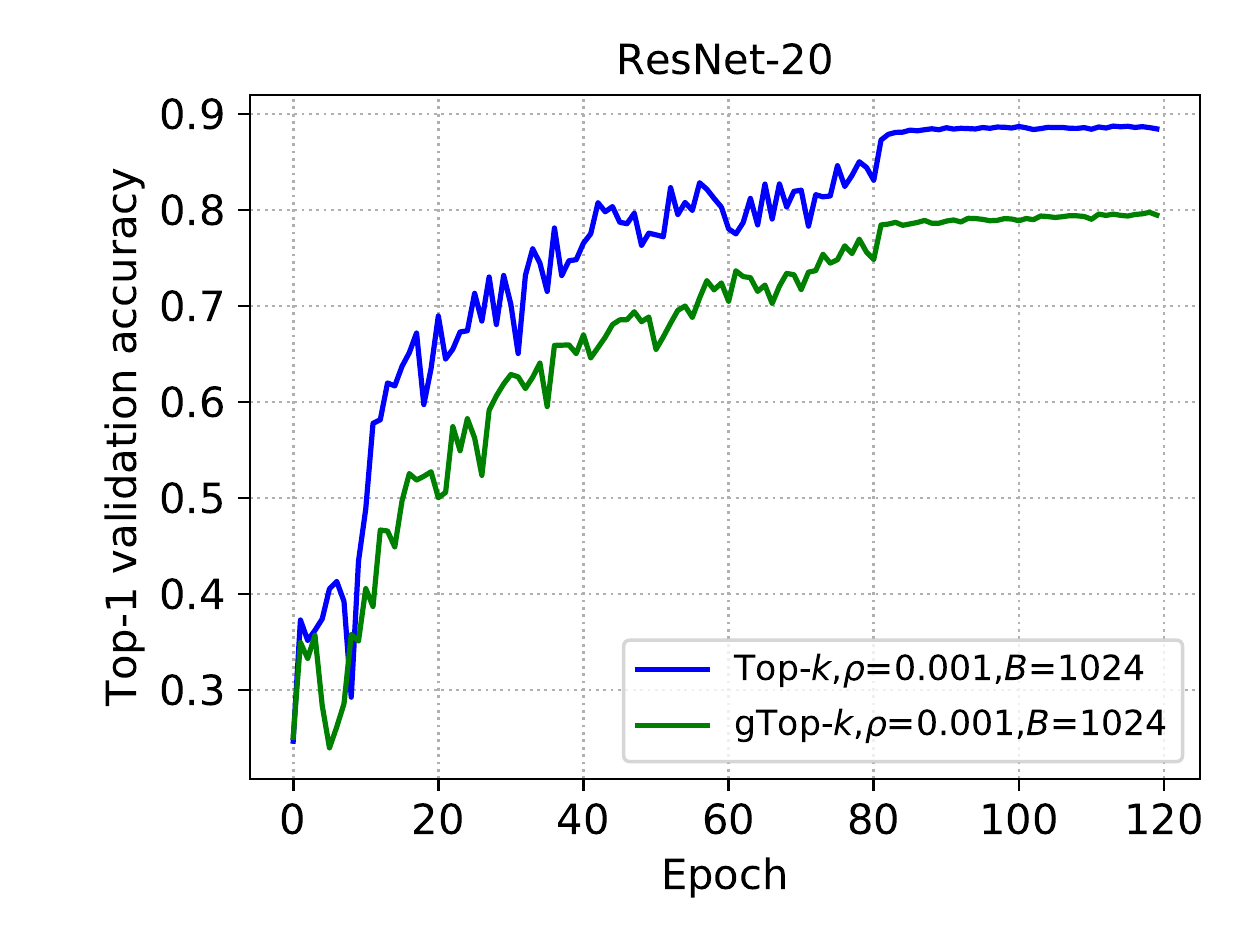}
	}\hspace{-5mm}
	\subfigure
	{
		\includegraphics[width=0.49\linewidth]{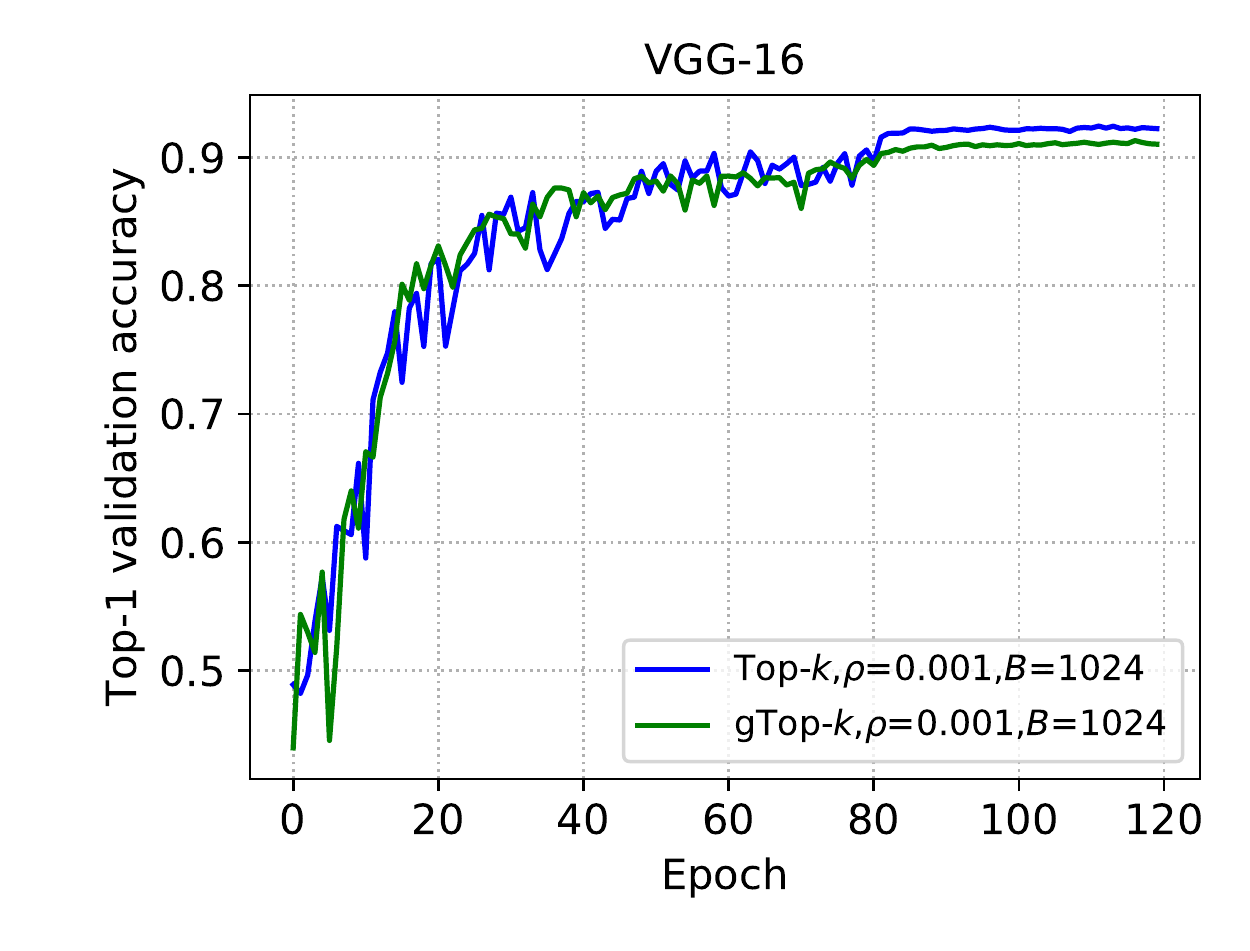}
	}
	\vspace{-18pt}
	\caption{The accuracy comparison between gTop-$k$ and Top-$k$ on ResNet-20 and VGG-16 with a mini-batch size of $1024$ and $P=32$.}
	\label{fig:acc1}
\end{figure}

\begin{figure}[!h]
	\centering
	\subfigure
	{
		\includegraphics[width=0.49\linewidth]{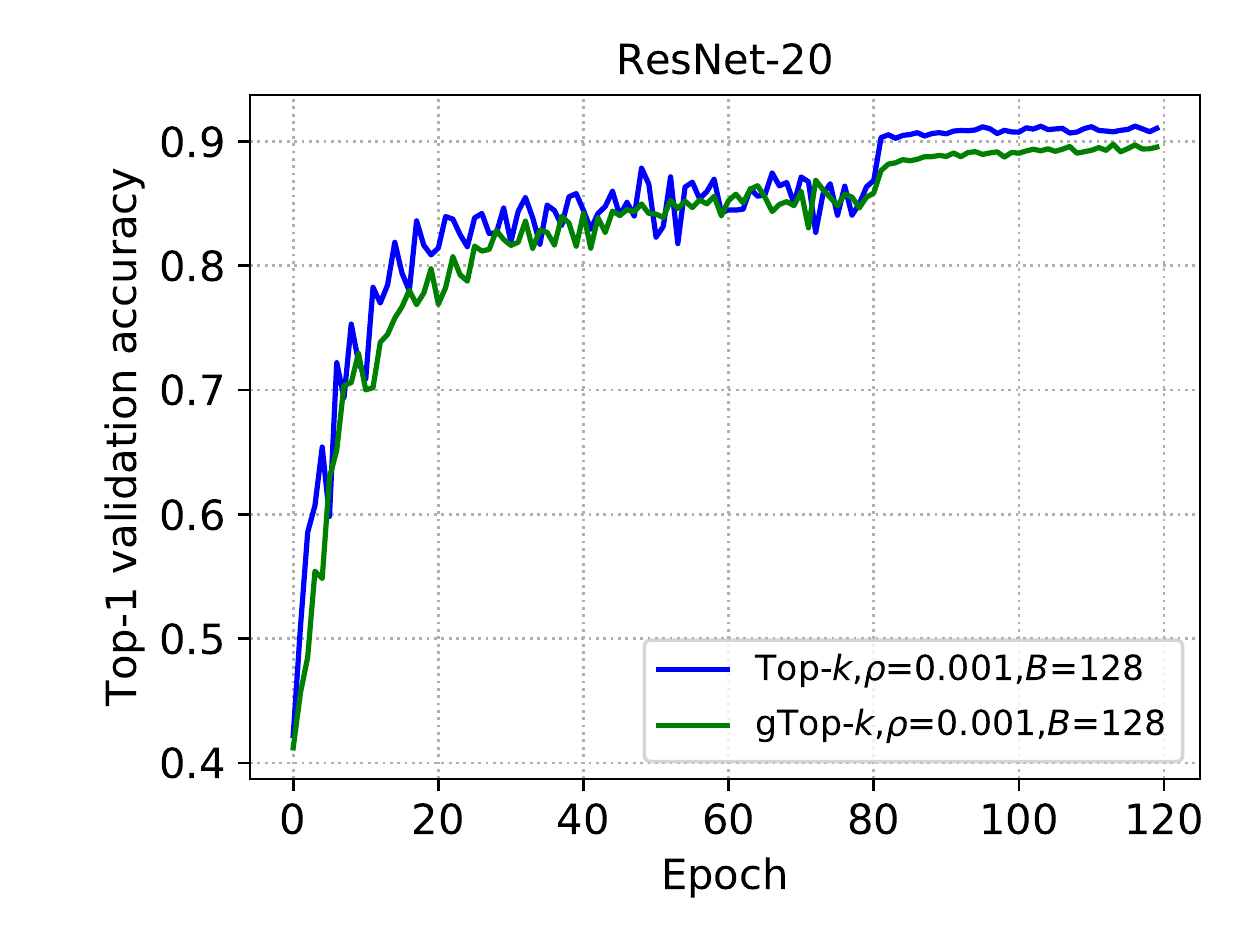}
	}\hspace{-5mm}
	\subfigure
	{
		\includegraphics[width=0.49\linewidth]{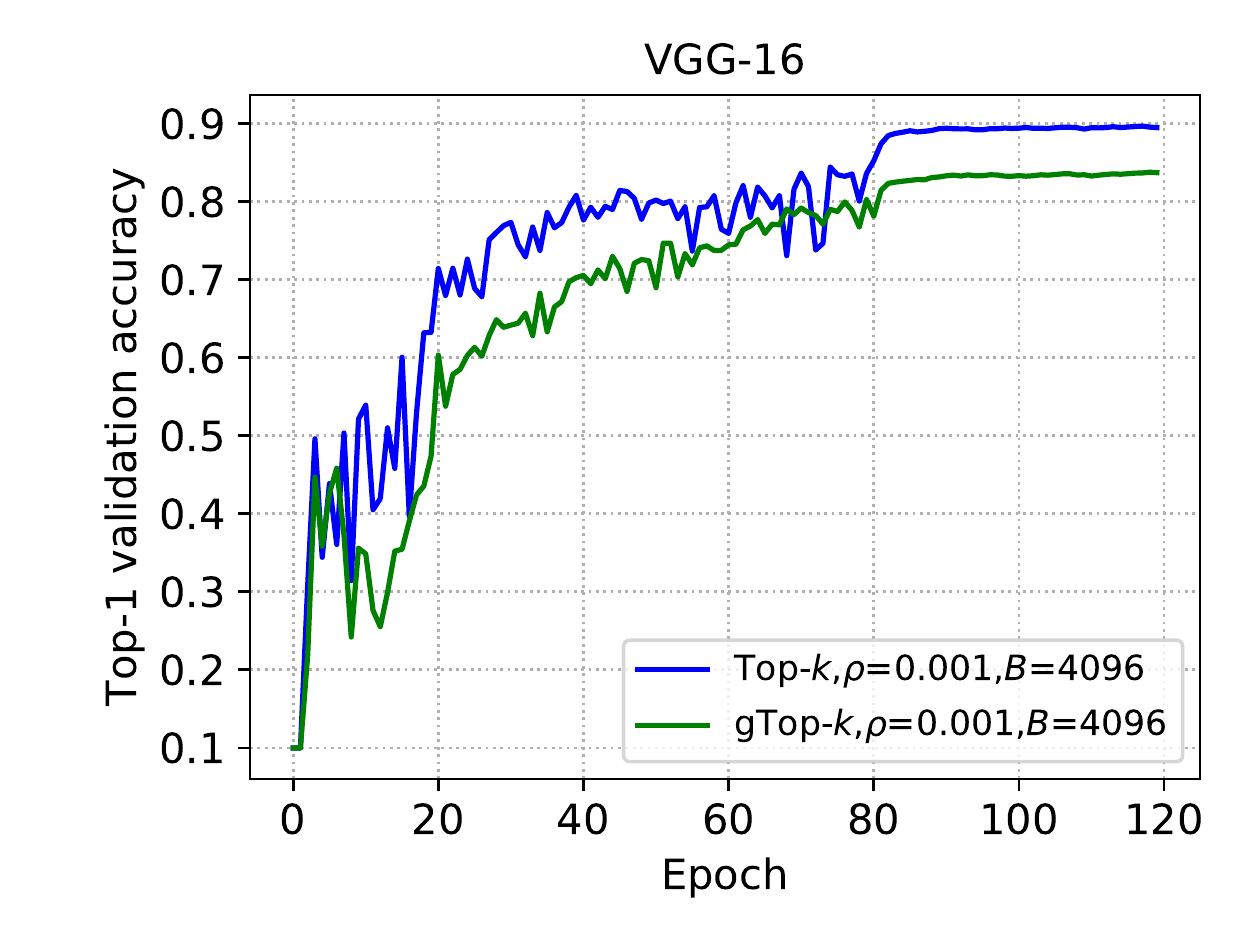}
	}
	\caption{The accuracy comparison between gTop-$k$ and Top-$k$ on ResNet-20 and VGG-16 with $P=32$ and changed mini-batch sizes.}
	\label{fig:acc2}
\end{figure}

The previous section illustrates that gTop-$k$ S-SGD achieves nearly consistent convergences with proper choose densities. However, compared with Top-$k$ S-SGD, gTop-$k$ S-SGD has some disadvantages which may result in poorer generalization performance when the total number of iterations ($N$) is relatively small. Assume that we set $k=0.001\times m$, $P=32$ and $B=b\times P=1024$ on the Cifar-10 data set, we have $N=5880$ with $120$ epochs. At each iteration, Top-$k$ S-SGD could make $k \times P$ weights be updated, while gTop-$k$ S-SGD updates $k$ weights. Therefore, gTop-$k$ S-SGD has only $6$ updates on some weights, while Top-$k$ S-SGD has about $192$. The top-1 validation accuracy of VGG-16 and ResNet-20 trained with gTop-$k$ S-SGD and Top-$k$ S-SGD are shown in Fig. \ref{fig:acc1}. It shows that ResNet-20 has \textasciitilde$9\%$ accuracy degradation with gTop-$k$ S-SGD, while VGG-16 has only \textasciitilde$1\%$ accuracy degradation with gTop-$k$ S-SGD. As a result, gTop-$k$ S-SGD requires more updates (by setting smaller mini-batch size) on ResNet-20 to achieve a higher accuracy, and it could also have higher accuracy degradation on VGG-16 by reducing the number of updates (by setting large mini-batch size). The comparison is shown in Fig. \ref{fig:acc2} with changed mini-batch sizes, which shows that gTop-$k$ S-SGD achieves closer accuracy (only \textasciitilde$0.5\%$ degradation) to Top-$k$ S-SGD with smaller mini-batch size on ResNet-20, and gTop-$k$ has \textasciitilde{$ 6\%$} accuracy degradation with a large mini-batch size on VGG-16.

\section{Related Work}\label{sec:related}
Gradient size reduction in communication is crucial for distributed synchronous SGD. Gradient quantization \cite{seide20141}\cite{wen2017terngrad} and sparsification are two main techniques. Gradient quantization can only achieve a maximum of $32\times$ reduction compared to the 32-bit gradients, while gradient sparsification is more aggressive than quantization. Gradient sparsification zero-outs a large proportion of gradients to reduce the communication size dramatically. Aji et al. \cite{aji2017sparse} and Chen et al. \cite{chen2017adacomp} empirically demonstrate that up to $99\%$ gradients are not needed to update the model at each iteration, which indicates that the gradients would be very sparse to convergent the model with accumulations of gradient residuals. Aji et al. \cite{aji2017sparse} use static threshold selection to determine $k$, while Chen et al. \cite{chen2017adacomp} propose a dynamic version. Lin et al. \cite{lin2017deep} further propose some optimization tricks (including the warmup strategy, momentum correction, and gradient clipping) to address the accuracy loss introduced by dropping a large number of gradients, and they show that Top-$k$ sparsification S-SGD can converge very close to S-SGD with dense gradients. The above techniques of quantization and sparsification can be combined to achieve a higher compression ratio of gradients with little accuracy loss. For example, Lin et al. \cite{lin2017deep} achieve up to 270$\times$ and 600$\times$ compression ratio without loss of accuracy. Researchers in \cite{renggli2018sparcml} have realized that efficient sparse AllReduce algorithms are non-trivial to implement, and they propose the AllGather solution. However, the AllGather method requires a linear increase cost with respect to the number of workers. Therefore, the AllGather could be inefficient when scaling to large-scale clusters.

\section{Conclusion and Future Work}\label{sec:conclusion}
In this paper, we first showed that the accumulating results from top-$k$ gradients can be further sparsified by choosing largest absolute gradients before updating the model, which has no much impact on the model convergence. Then we identified that the Top-$k$ sparsification is inefficient in averaging the gradients from all workers because the indices of the Top-$k$ gradients are not the same such that one should use the AllGather collective to collect all the Top-$k$ gradients and indices. The AllGather method for Top-$k$ aggregation (TopKAllReduce) is linear expensive to the number of workers (i.e., the communication complexity is $O(kP)$, where $P$ is the number of workers), so it would have very low scalability when scaling to large clusters. To this end, we proposed a global Top-$k$ (gTop-$k$) sparsification approach for S-SGD. The gradient aggregation algorithm based on gTop-$k$, named gTopKAllReduce, only requires a communication complexity of $O(k\log P)$. Experimental studies on various of deep neural networks including CNNs and RNNs were conducted to verify gTop-$k$ S-SGD has only slightly impact on the model convergence. The experiments conducted on the 32-GPU cluster inter-connected with 1 Gbps Ethernet showed that our proposed gTop-$k$ S-SGD has much higher scaling efficiency than S-SGD and Top-$k$ S-SGD.

Pipelining between computation and communication increases the scalability by optimally hiding the communication overheads \cite{shi2019mgwfbp}. In the future work, we would like to investigate layer-wise sparsification such that the communication overheads can be further overlapped by the computation tasks. 

\section{Acknowledgements}
The research was supported by Hong Kong RGC GRF grant HKBU 12200418. We acknowledge MassGrid.com for their support on providing the GPU cluster for experiments. 

\bibliographystyle{IEEEtran}
\bibliography{gTopK-ICDCS2019.bbl}

\end{document}